\documentclass[journal,twoside,web]{ieeecolor}
\pdfoutput=1 
\usepackage{tmi}
\usepackage[ieee,subfig,noadjustwidth]{definition}
\usepackage{seqsplit}

\newcommand{\footref}[1]{\textsuperscript{\ref{#1}}}

\definecolor{sblue}{HTML}{02BCD4}
\definecolor{sred}{HTML}{F44436}
\definecolor{spink}{HTML}{E91E62}
\definecolor{sgreen}{HTML}{8BC34A}
\definecolor{spurple}{HTML}{3F51B5}
\definecolor{slightgreen}{HTML}{CCDE3A}
\definecolor{sorange}{HTML}{FE9800}
\definecolor{sgolden}{HTML}{FFC108}

\newcommand{\ie}{i.e.}
\newcommand{\eg}{e.g.}
\newcommand{\etc}{etc}

\newcommand{\benchmark}{BrainGB\xspace}
\newcommand{\hejie}[1]{{\color{black}#1}}
\newcommand{\minor}[1]{{\color{black}#1}}

\definecolor{mygreen}{rgb}{0,0.6,0}
\definecolor{mymauve}{rgb}{0.58,0,0.82} 
\definecolor{ieee}{rgb}{0,0.541,0.855}

\begin{document}
\title{\benchmark: A Benchmark for Brain Network Analysis with Graph Neural Networks}

\author{Hejie Cui, Wei Dai, Yanqiao Zhu, Xuan Kan, Antonio Aodong Chen Gu \\ Joshua Lukemire, Liang Zhan, Lifang He, Ying Guo, Carl Yang
\thanks{H. Cui, W. Dai, X. Kan, A. C. Gu, and C. Yang are with the Department of Computer Science, Emory University.}
\thanks{Y. Zhu is with the Department of Computer Science, University of California, Los Angeles.}
\thanks{J. Lukemire and Y. Guo are with the Department of Biostatistics and Bioinformatics, Emory University.}
\thanks{L. Zhan is with the Department of Electrical and Computer Engineering, University of Pittsburgh.}
\thanks{L. He is with the Department of Computer Science and Engineering, Lehigh University.}
\thanks{Correspondence should be addressed to C. Yang (e-mail: j.carlyang@emory.edu).}
}

\maketitle

\begin{abstract}
Mapping the connectome of the human brain using structural or functional connectivity has become one of the most pervasive paradigms for neuroimaging analysis. Recently, Graph Neural Networks (GNNs) motivated from geometric deep learning have attracted broad interest due to their established power for modeling complex networked data. Despite their superior performance in many fields, there has not yet been a systematic study of how to design effective GNNs for brain network analysis. To bridge this gap, we present \benchmark, a benchmark for brain network analysis with GNNs. \benchmark standardizes the process by (1) summarizing brain network construction pipelines for both functional and structural neuroimaging modalities and (2) modularizing the implementation of GNN designs. We conduct extensive experiments on datasets across cohorts and modalities and recommend a set of general recipes for effective GNN designs on brain networks. To support open and reproducible research on GNN-based brain network analysis, we host the \benchmark website at \url{https://braingb.us} with models, tutorials, examples, as well as an out-of-box Python package. We hope that this work will provide useful empirical evidence and offer insights for future research in this novel and promising direction.
\end{abstract}


\section{Introduction}
\label{sec:intro}
\begin{figure*}
    \includegraphics[width=1\linewidth]{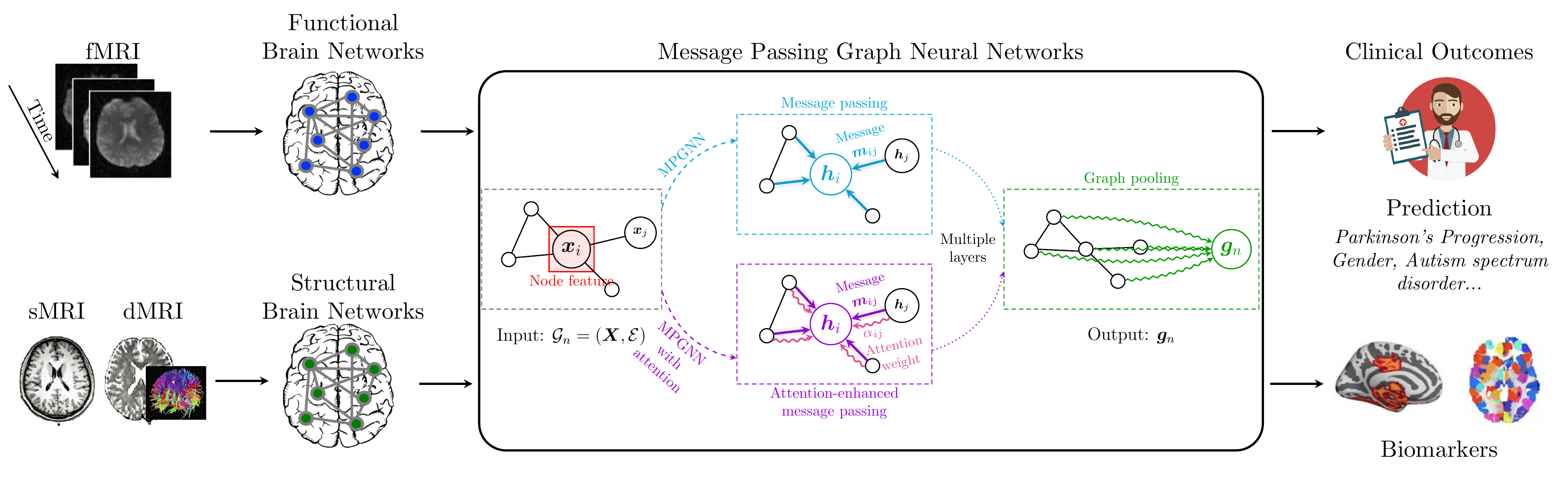}
    \caption{An overview of our \benchmark framework for brain network analysis with graph neural networks.}
    \label{fig:overview}
\end{figure*}
\IEEEPARstart{H}{uman} brains are at the center of complex neurobiological systems in which neurons, circuits, and subsystems interact to orchestrate behavior and cognition. Understanding the structures, functions, and mechanisms of human brains has been an intriguing pursuit for researchers with various goals, including neural system simulation, mental disorder therapy, as well as general artificial intelligence. Recent studies in neuroscience and brain imaging have reached the consensus that interactions between brain regions are key driving factors for neural development and disorder analysis \cite{li2020braingnn, farahani2019application}. Inspired by graph theory, brain networks composed of nodes and edges are developed to describe the interactions among brain regions. 

The human brain can be scanned through various medical imaging techniques, including Magnetic-Resonance Imaging (MRI), Electrogastrography (EGG), Positron Emission Tomography (PET), and so on. Among all these acquisitions, MRI data are the most widely used for brain analysis research. 
There are also different modalities of MRI data such as functional MRI (fMRI) and Diffusion Tensor Imaging (DTI), from which functional and structural brain networks can be constructed respectively. Specifically, the connectivity in functional brain networks describes correlations between time-series signals of brain regions, while the connectivity in structural brain networks models the physical connectivity between gray matter regions \cite{osipowicz2016functional}. Both functional and structural connections are widely acknowledged as valuable resources of information for brain investigation \cite{maglanoc2020multimodal, bullmore2009complex}. 

\hejie{Previous work on brain network analysis has studied shallow models based on graph theory \cite{bullmore2009complex, sporns2022graph} and tensor factorization \cite{Liu:2018ty,zhan2015boosting} extensively, which focuses on proposing neurobiologically insightful graph measures and approaches from the node, motif, and graph level to detect network communities or modules and identify central network elements. Methodological developments in graph research enable us to quantify more topological characteristics of complex systems, many of which have already been assessed in brain networks, such as modularity, hierarchy, centrality, and the distribution of network hubs. However, shallow modeling techniques can be inadequate for the sophisticated connectome structures of brain networks \cite{faskowitz2021edges}.} On the other hand, deep learning models have become extraordinarily popular in machine learning, achieving impressive performance on images \cite{DBLP:conf/iclr/DosovitskiyB0WZ21, radford2021learning}, videos \cite{arnab2021vivit}, and speech processing tasks \cite{DBLP:conf/interspeech/GulatiQCPZYHWZW20}. These regular data are represented in 1D/2D/3D Euclidean spaces and can be suitably handled by traditional Recurrent (RNNs) or Convolutional Neural Networks (CNNs). In contrast, the irregular structural and functional brain connectivity networks constructed from neuroimaging data are more complex due to their non-Euclidean characteristics. 
In recent years, Graph Neural Networks (GNNs) have attracted broad interest due to their established power for analyzing graph-structured data \cite{kipf2016semi, xu2019powerful, velivckovic2018graph}. Several pioneering deep models have been devised to predict brain diseases by learning graph structures of brain networks. 
For instance, \citet{li2020braingnn} propose BrainGNN to analyze fMRI data, where ROI-aware graph convolutional layers and ROI-selection pooling layers are designed for neurological biomarker prediction. \citet{kawahara2017brainnetcnn} design a CNN framework BrainNetCNN composed of edge-to-edge, edge-to-node, and node-to-graph convolutional filters that leverage the topological locality of structural brain networks. 
However, they mainly experiment with their proposed models on specific private datasets. Due to the ethical issue of human-related research, the datasets used are usually not publicly available and the details of imaging preprocessing are not disclosed, rendering the experiments irreproducible for other researchers.

To address the aforementioned limitations, there is an urgent need for a public benchmark platform to evaluate deep graph models for brain network analysis. 
However, it is non-trivial to integrate different components within a unified benchmarking platform. Current brain network analyses are typically composed of two steps. The first step is to construct brain networks from neuroimaging data. Then, in the second stage, the resulting brain connectivity between all node pairs is used to classify individuals or predict clinical outcomes. The difficulties in the initial stage are mostly due to restricted data accessibility and sophisticated brain imaging preprocessing and network construction pipelines that differ across cohorts and modalities. The difficulty of the second stage is to establish a standard evaluation pipeline based on fair experimental settings, metrics, and modular-designed baselines that can be easily validated and extended for future research. 

In this work, we propose Brain Graph Neural Network Benchmark (\benchmark)—a novel attempt to benchmark brain network analysis with GNNs to the best of our knowledge. The overview of \benchmark is demonstrated in Fig. \ref{fig:overview} and the main contributions are four-fold:
\begin{itemize}
    \item A \textit{unified}, \textit{modular}, \textit{scalable}, and \textit{reproducible} framework is established for brain network analysis with GNNs to facilitate reproducibility. It is designed to enable fair evaluation with accessible datasets, standard settings, and baselines to foster a collaborative environment within computational neuroscience and other related communities.
    \item We summarize the preprocessing and construction pipelines for both functional and structural brain networks to bridge the gap between the neuroimaging and deep learning community.
    \item We decompose the design space of interest for GNN-based brain network analysis into four modules: (1) node features, (b) message passing mechanisms, (c) attention mechanisms, and (d) pooling strategies. Different combinations based on these four dimensions are provided as baselines, and the framework can be easily extended to new variants.
    \item We conduct a variety of empirical studies and suggest a set of general recipes for effective GNN designs on brain networks, which could be a starting point for further studies.
\end{itemize}


To foster future research, we release \hejie{the} source code of BrainGB at \url{https://github.com/HennyJie/BrainGB} and provide an out-of-box package that can be installed directly, with detailed tutorials available on our hosted website at \url{https://braingb.us}. Preprocessing instructions and models are provided for standardized model evaluations.
We enable the community to collaboratively contribute by submitting their own custom models, and we will maintain a leaderboard to ensure such efforts will be recorded. 

\section{Preliminaries}
\label{sec:gnn}
\subsection{Brain Network Analysis}
Brain networks are complex graphs with anatomic Regions of Interest (ROIs) represented as nodes and connectivities between the ROIs as links \cite{murugesan2020brainnet}. In recent years, the analysis of brain networks has become increasingly important in neuroimaging studies to understand human brain organization across different groups of individuals \cite{su2020deep, satterthwaite2015linked, deco2011emerging, wang2019hierarchical, yu2019weighted}. Abundant findings in neuroscience research suggest that neural circuits are highly related to brain functions, with aberrations in these neural circuits being identified in diseased individuals \cite{insel2015brain, williams2016precision, li2016novel}. 

Formally, in the task of brain network analysis, the input is a brain network dataset $\mathcal{D} = \{\mathcal{G}_n, y_n\}_{n=1}^N$ consisting of $N$ subjects, where $\mathcal{G}_n=\{\mathcal{V}_n, \mathcal{E}_n\}$ represents the brain network of subject $n$ and $y_n$ is the subject's label of the prediction, such as neural diseases. In $\mathcal{D}$, the brain network $\mathcal{G}_n$ of every subject $n$ involves the same set of $M$ nodes defined by the ROIs on a specific brain parcellation, \ie, $\forall n, \mathcal{V}_n=\mathcal{V} = \{v_i\}_{i = 1}^M$. The difference across subjects lies in the edge connections $\mathcal{E}_n$ among $M$ brain regions, which are often represented by a weighted adjacency matrix $\bm{W}_n \in \mathbb{R}^{M \times M}$ describing the connection strengths between ROIs. The edge weights in $\bm{W}$ are real-valued and the edges are potentially dense and noisy. The model outputs a prediction $\hat{y}_n$ for each subject $n$, which can be further analyzed in terms of features and biomarkers. 

Given brain networks constructed from different modalities such as Diffusion Tensor Imaging (DTI) and functional Magnetic Resonance Imaging (fMRI) \cite{bullmore2009complex, zimmermann2018unique,hu2022multimodal}, effective analysis of the neural connectivities of different label groups (\eg, disease, gender) plays a pivotal role in understanding the biological structures and functions of the complex neural system, which can be helpful in the early diagnosis of neurological disorders and facilitate neuroscience research \cite{maartensson2018stability, yahata2016small, lindquist2008statistical, smith2012future,shi2016, dai2017predicting, higgins2019difference}. Previous models on brain networks are mostly shallow, such as graph kernels \cite{jie2016sub} and tensor factorization \cite{he2018boosted, liu2018multi}, which are unable to model the complex graph structures of the brain networks \cite{faskowitz2021edges}.

\subsection{Graph Neural Networks}
Graph Neural Networks (GNNs) have revolutionized the field of graph modeling and analysis for real-world networked data such as social networks \cite{kipf2016semi}, knowledge graphs \cite{schlichtkrull2018modeling}, protein or gene interaction networks \cite{xu2019powerful}, and recommendation systems \cite{wu2019session}. The advantage of GNNs is that they can combine node features and graph structures in an end-to-end fashion as needed for specific prediction tasks. 
A generic framework of GNN could be represented in two phases. In the first phase, it computes the representation $\bm{h}_i$ of each node $v_i\in\mathcal{V}_n$ by recursively aggregating messages from $v_i$'s multi-hop neighborhood, where $\bm{h}_i^0$ is initialized with node features. After getting the last-layer node representation $\bm{h}^{(L)}$, an extra pooling strategy is adopted to obtain the graph representation. Thereafter, a Multi-Layer Perceptron (MLP) can be applied to make predictions on the downstream tasks.

It is worth noting that brain networks are different from other real-world graphs such as social networks or knowledge graphs, due to (1) \hejie{the lack of useful initial node (ROI) features on brain networks represented by featureless graphs}, (2) the real-valued connection weights that can be both positive or negative, and (3) \hejie{the ROI identities and their orders are fixed across individual graph samples within the same dataset}. The design of GNN models should be customized to fit the unique nature of brain network data. Recently, there have been emerging efforts on GNN-based brain network analysis \cite{li2019graph, li2020braingnn, kawahara2017brainnetcnn, bessadok2021graph, cui2022interpretable, zhu2022joint, xuan2022fbnetgen,tang2022hierarchical2,tang2022hierarchical}. 
\hejie{However, these models are only tested on specific local datasets, mainly due to the convention in neuroscience that researchers are more used to developing methods that are applicable to their specific datasets and the regulatory restrictions that most brain imaging datasets are usually restrictively public, meaning that qualified researchers need to request access to the raw imaging data and preprocess them to obtain brain network data, but they are not allowed to release the preprocessed data afterwards. These challenges largely prohibit the methodology development in computational neuroscience research.}




\begin{figure*}
	\centering
    \includegraphics[width=1\linewidth]{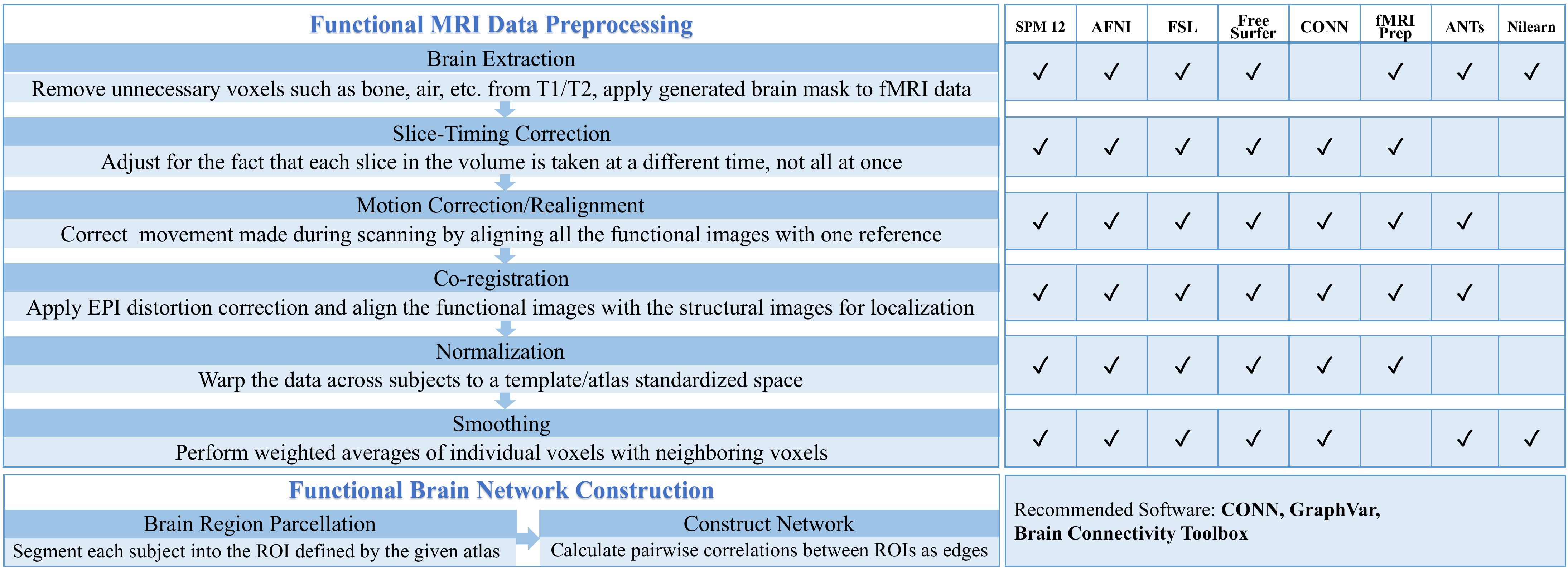}
	\caption{The framework of fMRI data preprocessing and functional brain network construction procedures, with recommended tools for each step shown on the right. The more commonly-used tools for the functional modality are placed at the front.}
	\label{fig:func_construct}
\end{figure*}
\section{Brain Network Dataset Construction}
\label{sec:datasets}
\begin{figure*}
	\centering
    \includegraphics[width=1\linewidth]{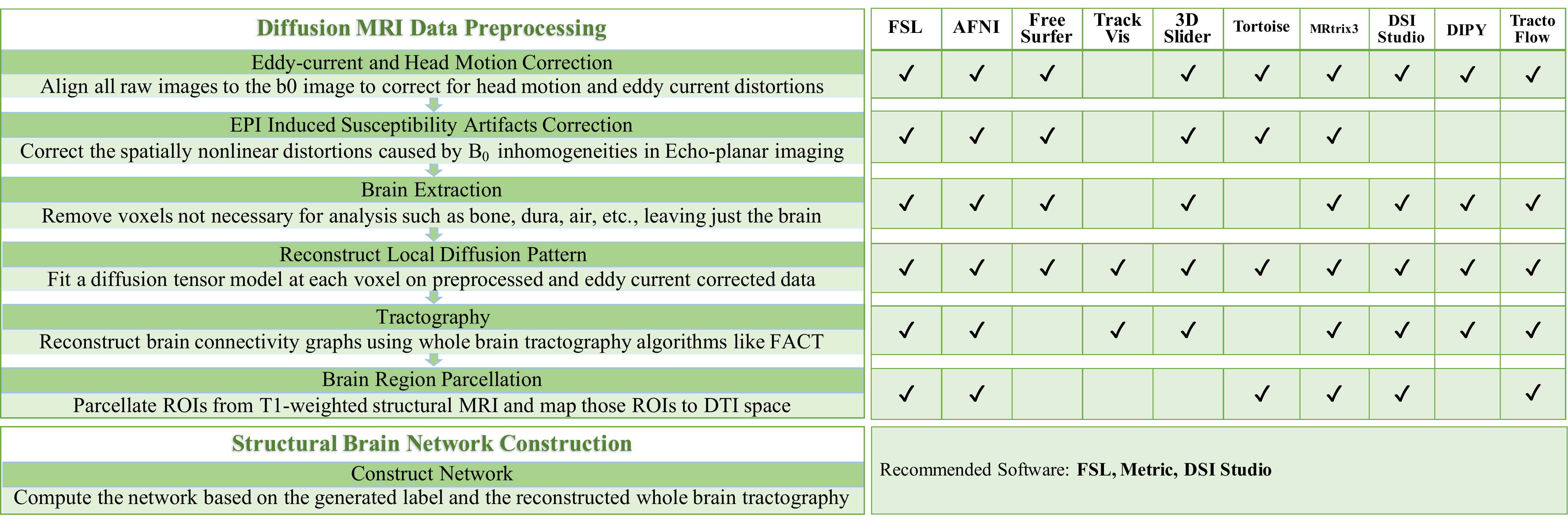}
	\caption{The framework of dMRI data preprocessing and structural brain network construction procedures, with recommended tools for each step shown on the right. The more commonly-used tools for the structural modality are placed at the front.}
	\label{fig:struc_construct}
\end{figure*}
\subsection{Background: Diverse Modalities of Brain Imaging}
Models of the human brain as a complex network have attracted increasing attention due to their potential for helping understand human cognition and neurological disorders. In practice, human brain data can be acquired through various scanning techniques \cite{sarraf2016functional}, such as Magnetic-Resonance Imaging (MRI), Electroencephalography (EEG) and Magnetoencephalography (MEG), Positron Emission Tomography (PET), Single-Photon Emission Computed Tomography (SPECT), and X-ray Computed Tomography (CT). Among them, \hejie{MRI is one of the most widely used techniques in brain research and clinical practice}, due to its large range of available tissue contrast, detailed anatomical visualization, and high sensitivity to abnormalities \cite{DBLP:journals/mda/BernsteinAKSSG18}. 

\subsubsection{MRI Data}
In this paper, we focus on MRI-derived brain networks. Specifically, for different modalities of MRI data, we can reconstruct different types of brain networks. Functional MRI (fMRI) is one of the most popular modalities for investigating brain function and organization \citep{GANIS2002493,lindquist2008statistical,smith2012future} by detecting changes in blood oxygenation and blood flow that occur in response to neural activity. Diffusion-weighted MRI (dMRI), on the other hand, can enable inference about the underlying connection structure in the brain's white matter by recording the diffusion trajectory of molecules (usually water). fMRI focuses on functional activity, while dMRI presents brain structural information from different perspectives. Specifically, two types of brain networks, functional and structural, can be constructed from the aforementioned modalities by following different connectivity generation paradigms \cite{DBLP:journals/cin/LangTKSP12}.

\subsubsection{Challenges in MRI Preprocessings}
The raw MRI data collected from scanners is not directly usable for brain network construction or imaging analysis. A complicated preprocessing pipeline is necessary to remove unwanted artifacts, transform the data into a standard format, and perform structure discovery. Although there are several widely-used neuroimaging data preprocessing tools, such as SPM\footnote{\label{spm}\url{https://www.fil.ion.ucl.ac.uk/spm/software/spm12}}, AFNI\footnote{\label{afni}\url{https://afni.nimh.nih.gov}} and FSL\footnote{\label{fsl}\url{https://fsl.fmrib.ox.ac.uk/fsl/fslwiki/FSL}}, each of them still needs considerable training and learning efforts. Moreover, the functionality of these software varies, and for dMRI, no one software contains all the necessary preprocessing capabilities. In addition, many neuroimaging datasets cannot be made public due to privacy or ethical concerns. Due to the variety of preprocessing approaches and issues with making data publically available, there are difficulties in reproducibility in neuroimaging studies. Additionally, the preprocessing steps are distinctive across modalities. All these challenges make it difficult for deep learning researchers with little knowledge in medical imaging processing to get into the field. 

\subsection{Brain Network Construction from Raw Data}


\hejie{
In this section, we provide a general overview of the standard preprocessing pipelines for the construction of brain networks of different modalities. Due to the regulation restrictions for direct sharing of the brain network data, we provide two complete pipelines, one for functional brain networks (ABCD\footnote{\url{https://nda.nih.gov/abcd}\label{foot:abcd}} specifically) and one for structural brain networks (PPMI\footnote{\url{https://www.ppmi-info.org}\label{foot:ppmi}} specifically), with step-by-step commands and parameter settings on our hosted website for public access\footnote{\url{https://braingb.us/preprocessing}\label{website}}}.

\subsubsection{Functional Brain Network Construction}

The left side of Fig. \ref{fig:func_construct} shows a standard preprocessing procedure for functional brain imaging, with the corresponding commonly-used toolboxes (\ie, SPM12\footref{spm}, AFNI\footref{afni}, FSL\footref{fsl}, FreeSurfer\footnote{\label{freesurfer}\url{https://surfer.nmr.mgh.harvard.edu}}, CONN\footnote{\url{https://web.conn-toolbox.org/home}}, fMRI Prep\footnote{\url{https://fmriprep.org/en/stable/index.html}}, ANTs\footnote{\label{ants}\url{http://stnava.github.io/ANTs}}, Nilearn\footnote{\url{https://nilearn.github.io/stable/index.html}}) shown on the right side. Note that \hejie{each step in the preprocessing and network construction pipeline needs quality control by the experts, and the specific order of preprocessing steps} may change slightly based on the acquisition conditions of the dataset. Some representative functional neuroimaging datasets in literature to facilitate scientific research include ADHD 200 \cite{bellec2017neuro}, ADNI (fMRI part) \cite{petersen2010alzheimer}, HCP 900\cite{van2012human}, ABIDE \cite{di2014autism}, \etc. 

To measure functional connectivity, some preprocessing of the fMRI time series is often performed including detrending, demeaning, and whitening fMRI BOLD time series at each voxel \citep{wang2016efficient}. To construct the brain networks, a brain atlas or a set of Regions of Interest (ROI) are selected to define the nodes. Then, the representative fMRI BOLD series from each node are obtained by either averaging or performing \hejie{Singular Value Decomposition (SVD)} on the time series from all the voxels within the node. Various measures have been proposed for assessing brain connectivity between pairs of nodes. One of the simplest and most frequently used methods in the neuroimaging community is via pairwise correlations between BOLD time courses from two 
ROIs. Other methods include partial correlations  \citep{wang2016efficient}, mutual information, coherence, Granger causality \citep{smith2011network}. After selecting the Functional Connectivity (FC) measure, one can evaluate the strength of connectivity between each pair of ROIs. Often, some transformation, such as the Fisher's transformation, is performed to transform the original FC measures to improve their distribution properties.  
The transformed FC measures can then be utilized for the subsequent analysis of functional brain networks.

\hejie{To facilitate public testing, we take Adolescent Brain Cognitive Development Study (ABCD) as an example and provide a step-by-step instruction for functional brain network construction on our hosted BrainGB website\footref{website}. \minor{The ABCD-HCP BIDS\footnote{\url{https://github.com/DCAN-Labs/abcd-hcp-pipeline}\label{foot:abcdpip}} pipeline is used to preprocess the data. In brief, anatomical preprocessing included normalization, co-registration, segmentation, and brain extraction. Functional data preprocessing included slice-time correction, motion correction, distortion correction, co-registration, normalization, and spatial smoothing. Brain parcellation schemes were then applied to the functional data to obtain time courses for each ROI, and Pearson correlation was used to construct brain networks representing the connectivity between ROIs.} } 
\subsubsection{Structural Brain Network Construction}
Structural brain networks provide a systematic perspective for studying the anatomical and physiological organization of human brains and help to understand how brain structure influences function. 
Some representative neuroimaging studies include diffusion MRI data are PPMI \cite{marek2011parkinson}, ADNI \cite{petersen2010alzheimer}, HCP \cite{van2012human}, AIBL \cite{ellis2009australian}, OASIS \cite{ellis2009australian}, \etc. The commonly-used toolboxes for dMRI include FSL\footref{fsl}, AFNI\footref{afni}, FreeSurfer\footref{freesurfer}, TrackVis\footnote{\url{http://trackvis.org}}, 3D Slicer\footnote{\url{https://www.slicer.org}}, Tortoise\footnote{\url{https://tortoise.nibib.nih.gov}}, MRtrix3\footnote{\url{https://www.mrtrix.org}}, DSI Studio\footnote{\url{https://dsi-studio.labsolver.org}}.

The left side of Fig. \ref{fig:struc_construct} summarizes the pipeline for reconstructing the structural brain network.
Preprocessing steps for the dMRI data include removal of eddy current-induced distortions, brain extraction, and co-registration between diffusion and structural images. Next, some modeling strategies are applied to reconstruct the local diffusion patterns. Commonly adopted models include the DTI modeling, which fits a tensor model or muti-tensor model \cite{leow2009tensor} to capture the local diffusion patterns, and the Ball and Sticks model \citep{behrens2003characterization}. After reconstructing the local diffusion patterns, a tractography algorithm is performed to computationally reconstruct fiber tract connections between brain regions. Commonly-used algorithms include the deterministic tractography \citep{basser2000vivo} and the probabilistic tractography \citep{behrens2007probabilistic}. The deterministic tractography connects neighboring voxels from seed regions based on the major direction of the DTI tensor. The probabilistic tractography involves first estimating fiber orientation and its uncertainty at each voxel and building a diffusion path probability map based on the estimated orientation and uncertainty. While deterministic tractography is a more computationally efficient approach to reconstruct major fiber bundles in the brain, probabilistic tractography has become more popular because it is more robust to noise and allows tractography to progress beyond uncertain regions by taking into account uncertainty in fiber orientations at each voxel \cite{zhan2015comparison}. To construct the structural network, the structure connectivity for each node pair is calculated based on the empirical probability of fiber tracts connecting the two regions. \hejie{Note that each step of network construction ideally needs quality control from experts.}

\hejie{Similarly to functional brain network construction, we take PPMI as an example and provide an instruction pipeline for structural brain network construction on our hosted BrainGB website\footref{website}. Specifically, the Diffusion Toolkit from TrackVis is used to reconstruct local diffusion patterns and tractography. The brain region parcellation is completed with both FSL and Freesurfer. Then local diffusion pattern reconstruction and the network computation are further performed by calculating the number of fibers within each ROI after removing the false positive ones.} 













\subsection{Discussions}
In addition to the mainstream methods of constructing connections in brain networks discussed above, there are also other ways to construct different types of edges. For example, directional connectivity that characterizes effective interactions for fMRI \cite{deshpande2020multi}; hybrid functional brain networks where different orders of relationships can be sensitive to different levels of signal changes \cite{zhu2019hybrid}; and dynamic functional brain networks which include derivatives of windowed functional network connectivity in the identification of reoccurring states of connectivity \cite{espinoza2019characterizing, deshpande2020multi}. 
Apart from fMRI and DTI, the most commonly used modalities to construct functional and structural brain networks, other neuroimaging modalities have also been explored in literature, such as metabolic brain network constructed from PET imaging \cite{kuang2019metabolic}, functional brain network constructed from EEG signals \cite{joudaki2012eeg}, \etc. 
Recent studies have shown that the combination of both functional and structural neuroimaging modalities can be more effective than using only a single one, which can exploit complementary information across different modalities \cite{maglanoc2020multimodal, calhoun2016multimodal}.

\section{GNN Baselines for Brain Network Analysis}
\label{baselines}
The process of applying GNNs to brain networks starts from initialization of the ROI features, followed by the forward pass which includes two phases, message passing, and pooling. The learned graph-level representation then can be utilized for brain disease analysis. 
In the machine learning domain, the rapid evolution of GNNs has led to a growing number of new architectures. Specifically for GNNs on brain network analysis, we decompose the design space of interest for basic message passing GNNs into four modules: node feature construction, message passing, attention enhanced message passing, and pooling strategies. An illustration of these modules is shown in the middle of Fig. \ref{fig:overview}. 

\subsection{Node Feature Construction}
\hejie{In neuroscience analysis, researchers mostly focus on brain connectivity represented by a featureless graph.} To apply GNNs on non-attributed brain networks, researchers in the graph machine learning domain have studied several practical methods to initialize node features \cite{cui2021positional, duong2019node}. In this paper, we focus on the following node features that can be categorized as positional or structural:
\begin{itemize}
    \item \textit{Identity}: A unique one-hot feature vector is initialized for each node \cite{errica_fair_2020, you2019position}. \hejie{By giving each ROI in the brain network a unique high-dimensional vector, this identity node feature allows the GNN model to learn the relative positions of the nodes by memorizing their k-hop neighbors. They are essentially the same as random initialization considering the parameters in the first linear layer of the GNN are randomly initialized.}
    \item \textit{Eigen}: Eigen decomposition is performed on the weighted matrix \hejie{describing the connection strengths between ROIs} and then the top \textit{k} eigenvectors are used to generate a \textit{k}-dimensional feature vector for each node \cite{DBLP:journals/corr/abs-2010-13993, chaudhuri2012spectral, DBLP:conf/nips/ZhangR18}. The optimal value of \textit{k} is decided by grid search. This feature is essentially dimension reduction and targets at grouping brain regions with respect to their positions, with global graph information condensed into a low-dimensional representation.
    \item \textit{Degree}: The degree value of each node is obtained as a one-dimensional vector as the node feature. This feature captures structural information of brain regions, meaning that neighborhood structural similarity of two regions will be partially recorded in the initialized node features.
    \item \textit{Degree profile}: This method takes advantages of existing local statistical measures on degree profiles \cite{Cai:2018te}, where each feature \(\bm{x}_i\) of node \(v_i\) on graph $\mathcal{G}_n$ is computed as
    \begin{equation}
        \begin{aligned}
            \bm{x}_i = [\deg(v_i)\;\|\; \min(\mathcal{D}_i) \;\|\; \max(\mathcal{D}_i) \\
            \;\|\; \operatorname{mean}(\mathcal{D}_i) \;\|\; \operatorname{std}(\mathcal{D}_i) ],
        \end{aligned}
    \end{equation}
    where \(\mathcal{D}_i = \{\deg(v_i) \mid (i,j) \in \mathcal{E}_{n}\}\) \hejie{describes the degree values of} node \(v_i\)'s one-hop neighborhood and \( \;\|\; \) denotes concatenation. 
    \item \textit{Connection profile}: The corresponding row for each node in the edge weight matrix is utilized as the initial node feature, which contains connections with respect to all other nodes in the brain network. This feature aligns with the common practice of using pairwise connections to perform brain parcellation. Also, it reflects the whole picture of connection information in the brain network.
\end{itemize}

\subsection{Message Passing Mechanisms}
\label{mp}
The power of most GNNs to learn structures lies in their message passing schemes, where the node representation is updated iteratively by aggregating neighbor features through local connections. 
\hejie{In each layer $l$, the node representation $\bm{h}_{i}^{l}$ is updated through two steps, namely message passing and update respectively.
In the message passing step (Eq. \ref{eq:message}), each node $v_i$ receives messages from all its neighbors, and then all the messages are aggregated with a sum function:
\begin{align}
\bm{m}_{i}^{l}=\sum_{j \in \mathcal{N}_i} \bm{m}_{ij}&=\sum_{j \in \mathcal{N}_i} M_{l}\left(\bm{h}_{i}^{l}, \bm{h}_{j}^{l}, w_{ij}\right),
\label{eq:message}
\end{align}
where $\mathcal{N}_{i}$ denotes the neighbors of node $v_{i}$ in graph $\mathcal{G}$, $w_{ij}$ represents the edge weights between node $v_i$ and $v_j$, $M_l$ is the message function.
In the update step (Eq. \ref{eq:update}), the embedding of each node is updated based on the aggregated messages from Eq. \ref{eq:message} and optionally the previous embedding of node $v_i$, where the update function can be arbitrary differentiable functions (e.g., concat the aggregated message with the previous node embedding and then pass them into a learnable linear layer). }
\begin{align}
\bm{h}_{i}^{l+1}&=U_{l}\left(\bm{h}_{i}^{l}, \bm{m}_{i}^{l}\right),
\label{eq:update}
\end{align}
where $U_l$ stands for the update function and the number of running steps $L$ is defined by the number of GNN layers. 
The message passing mechanism can leverage both permutation equivariance and inductive bias towards learning local structures and achieve good generalization on new graphs. For brain networks, whether incorporating connections into the message function is beneficial for graph-level prediction tasks remains to be investigated. In this paper, we discuss the influence of different \minor{message function} \hejie{$M_{l}$} designs including:
\begin{itemize}
\item \textit{Edge weighted}: The message $\bm{m}_{ij}$ passed from node $v_j$ to node $v_i$ is calculated as the representation of node $v_j$ weighted by the corresponding edge weight $w_{ij}$, that is 
\begin{align}\label{eq:weighted_sum}
\bm{m}_{ij} = \bm{h}_{j} \cdot w_{ij}.
\end{align}
This is the standard message passing implementation in Graph Convolutional Network (GCN) \cite{kipf2016semi} when $w_{ij}=\nicefrac{1}{N_i}$. 
With this message vector design, the update of each brain region representation is influenced by its neighbor regions weighted by the connection strength between them. 
\item \textit{Bin concat}: In this scheme, we map the edge $w_{ij}$ into one of the \minor{equally split $T$ buckets} based on its weight value. \hejie{Each bucket corresponds to a learnable representation \minor{$\bm{b}_{t}$, $t= \left\{1 \ldots T\right\}  $}. The total bucket number encompassing the entire value range of edge weights is determined by grid search and the representation dimension of each bin is set to the same as node features.
Specifically, given the number of buckets is \minor{$T$}, we first rank all the edge weights and then divide them into the \minor{equally divided $T$} buckets from the lowest to the highest. All edges in the same bucket will be mapped to the same learnable vector $\bm{b}_{t}$, so region connections with similar strength are binned together. In our experiment, we simply select from [5, 10, 15, 20] as the possible number of buckets for grid search, which is a common practice in machine learning for hyperparameter tuning.} The message $\bm{m}_j$ passed from node $v_j$ to node $v_i$ is calculated as the concatenation of the representation of node $v_j$ and \minor{its corresponding bucket representation $\bm{b}_{t}$} followed by an MLP, 
\begin{align}
\bm{m}_{ij} = \operatorname{MLP}(\bm{h}_{j} \;\Vert\; \bm{b}_{t}).
\end{align}
The usage of bins helps to clusters region connections with similar strengths. By concatenating with the unique neighbor node representation, this message captures both common and peculiar characteristics of each neighbor.
\item \textit{Edge weight concat}: 
The message $\bm{m}_{ij}$ passed from node $v_j$ to node $v_i$ is represented as the concatenation of the representation of node $v_j$ and \hejie{the scaled edge weight $d \cdot w_{ij}$}, followed by a MLP, 
\begin{align}
\bm{m}_{ij} &= \hejie{\operatorname{MLP}(\bm{h}_{j} \;\Vert\; d \cdot w_{ij})},
\end{align}
\minor{where $d$ is a constant equals to the dimension number of node features.}
\hejie{The motivation behind edge weight scaling is to increase the influence of edge features to the same scale as node features. Compared with bin concat where edges with weight values in the same bin interval share the same initial edge representation, directly concatenating the scaled edge weights as the edge representations can retain the original edge information, therefore reserving more uniqueness on the pairwise connection when performing the aggregation from neighboring brain regions.}
\item \textit{Node edge concat}: To investigate the influence of preserving the brain region representation from the last time step while iterative updating the new representation, we design a message $\bm{m}_j$ as the concatenation of both embeddings of node $v_i$, $v_i$ and the edge weight $w_{ij}$ between them, followed by a MLP, that is
\begin{align}\label{eq:edge_node_concat}
\bm{m}_{ij} = \operatorname{MLP}(\bm{h}_{i} \;\Vert\; \bm{h}_{j} \;\Vert\; w_{ij}).
\end{align}
In this paradigm, every message passed from the local neighbors of each central node is reinforced with its representation from the last time step. This design may alleviate the over-smoothing problem of GNNs, where the feature distance between all nodes becomes too close and not distinguishable after layers of convolutions.
\item \textit{Node concat}: Since the effect of involving connection weights into message passing is still unknown, we also include another message $\bm{m}_{ij}$ similar to \textit{node edge concat} but without the concatenation of edge weights, where
\begin{align}\label{eq:node_concat}
\bm{m}_{ij} = \operatorname{MLP}(\bm{h}_{i} \;\Vert\; \bm{h}_{j}).
\end{align}
\end{itemize}

\subsection{Attention-Enhanced Message Passing}
Attention is arguably one of the most important mechanisms in modern deep learning \cite{vaswani2017attention, niu2021review}. It is inspired by human cognitive systems that tend to selectively concentrate on the important parts as needed when processing large amounts of information. Various fields in deep learning communities such as natural language processing \cite{devlin-etal-2019-bert} and computer vision \cite{guo2021attention_survey} have widely benefited from attention mechanisms in terms of model efficiency and accuracy. 
The attention mechanism can also be used to enhance the message passing scheme of GNNs, while also providing interpretations over the edge importance.

Specifically in brain network analysis, by utilizing the attention-enhanced version of message passing, the model updates the brain region representation in a data-driven way, where adjustable attention weights from each local neighbor perform as an additional influence factor besides the neural signals represented by edge weights. It is worth noting that the traditional designs of graph attention mechanisms on general graphs usually do not take the edge attributes (\ie, connection weights in the brain network scenario) into consideration. However, for brain networks, the correlation between two regions contains meaningful biomedical information and might be helpful for graph-level tasks. In this paper, we design several attention-enhanced message passing mechanisms including: 
\begin{itemize}
\item \textit{Attention weighted}: This is the original GAT  \cite{velivckovic2018graph} on general graphs without involving edge attributes. The message from node $v_j$ to $v_i$ is weighted by the corresponding attention score  $\alpha_{ij}$ as 
\begin{align}
    \bm{m}_{ij} = \bm{h}_{j} \cdot \alpha_{ij}.
\end{align}
The $\alpha_{ij}$ is calculated from a single-layer feed-forward neural network parameterized by a weight vector $\mathbf{a}$, followed by the $\operatorname{LeakyReLU}$ nonlinearity $\sigma$,
\begin{align}
\alpha_{ij}=\frac{\exp \left(\sigma\left(\bm{a}^{\top}\left[\bm{\Theta} \bm{x}_{i} \;\|\; \bm{\Theta} \bm{x}_{j}\right]\right)\right)}{\sum_{k \in \mathcal{N}(i) \cup\{i\}} \exp \left(\sigma\left(\bm{a}^{\top}\left[\bm{\Theta} \bm{x}_{i} \;\|\; \bm{\Theta} \bm{x}_{k}\right]\right)\right)},
\label{attn}
\end{align}
where $\bm{\Theta}$ is a learnable linear transformation matrix.
\item \textit{Edge weighted w/ attn}: This is the attention-enhanced version of \textit{edge weighted} message passing in Eq. \ref{eq:weighted_sum}. The message from $v_j$ to $v_i$ is obtained as the multiplication of node $v_j$'s representation $\bm{h}_j$, the edge weight $w_{ij}$ and the attention score $\alpha_{ij}$ in Eq. \ref{attn},
\begin{align}
    \bm{m}_{ij} = \bm{h}_{j} \cdot \alpha_{ij} \cdot w_{ij}.
\end{align}
\item \textit{Attention edge sum}: This is another version of attention-enhanced \textit{edge weighted} (Eq. \ref{eq:weighted_sum}) message passing. The edge weight $w_{ij}$ and the attention score $\alpha_{ij}$ are first summed, then used as the impact factor on the node embedding $\bm{h}_j$, 
\begin{align}
    \bm{m}_{ij} = \bm{h}_{j} \cdot (\alpha_{ij} + w_{ij}).
\end{align}

\item \textit{Node edge concat w/ attn}: This is the attention-enhanced version of \textit{node edge concat} (Eq. \ref{eq:edge_node_concat}) message passing, where the attention score $\alpha_{ij}$ (Eq. \ref{attn}) between node $v_i$ and $v_j$ is multiplied on the node representation $\bm{h}_{j}$ before concatenation, followed by a MLP,
\begin{align}\label{Node edge concat w/ attn}
    \bm{m}_{ij} &= \operatorname{MLP}(\bm{h}_{i} \;\Vert\; (\bm{h}_{j} \cdot \alpha_{ij}) \;\Vert\; w_{ij}).
\end{align}
\item \textit{Node concat w/ attn}: This design corresponds to the attention-enhanced version of \textit{node concat} (Eq. \ref{eq:node_concat}) message passing, where the attention score $\alpha_{ij}$ (Eq. \ref{attn}) between node $v_i$ and node $v_j$ is multiplied on the node representation $\bm{h}_{j}$ before concatenation, followed by a MLP,
\begin{align}\label{Node concat w/ attn}
    \bm{m}_{ij} &= \operatorname{MLP}(\bm{h}_{i} \;\Vert\; (\bm{h}_{j} \cdot \alpha_{ij})).
\end{align}
\end{itemize}

\subsection{Pooling Strategies}
In the second phase of GNNs, a feature vector for the whole graph $\bm{g}_{n}$ is computed using the pooling strategy $R$, where
\begin{align}
    \bm{g}_{n} = R\left(\left\{\bm{h}_{\hejie{k}} \mid v_\hejie{k} \in \mathcal{G}_n\right\}\right).
\end{align}
The pooling function $R$ operates on the set of node vectors and is invariant to permutations of the node vectors. In this paper, we cover three basic global pooling operators \cite{grattarola2021understanding, mesquita2020rethinking}: 
\begin{itemize}
    \item \textit{Mean pooling}: The graph-level representation is obtained by averaging node features. For each single graph $\mathcal{G}_n$, the graph-level representation is computed as 
    \begin{align}
    \bm{g}_{n}=\frac{1}{M} \sum_{k=1}^{M} \bm{h}_{k}.
    \end{align}
    \item \textit{Sum pooling}: The graph-level representation is obtained by summing up all node features. For each single graph $\mathcal{G}_n$, the graph-level representation is computed as
    \begin{align}
    \bm{g}_{n}= \sum_{k=1}^{M} \bm{h}_{k}.
    \end{align}
    \item \textit{Concat pooling}: The graph-level representation is obtained by concatenating node features of all nodes contained in the graph. For each single graph $\mathcal{G}_n$, the graph-level representation is computed as
    \begin{align}\label{concat}
    \bm{g}_{n}= \Vert_{k=1}^{M}\; \bm{h}_{\hejie{k}} = \bm{h}_{1} \;\Vert\; \bm{h}_{2} \;\Vert\; \dots \;\Vert\; \bm{h}_{\hejie{M}}.
    \end{align}
\end{itemize}
\hejie{Note that there are also other complex pooling strategies such as hierarchical pooling \cite{ying2018hierarchical}, learnable pooling \cite{gopinath2020learnable} and clustering readout \cite{kan2022transformer}, which are usually viewed as independent GNN architecture designs that are not defined based on combinative modules. Here we include the representative method of DiffPool \cite{ying2018hierarchical} to provide a view of the comparison between basic and more complex pooling methods.}
\section{Experimental Analysis and Insights}

\begin{table*}
\centering
\scriptsize
\caption{\hejie{Dataset summarization.}}
\begin{tabular}{ccccccc}
\toprule

 Dataset & Modality & \# Samples & Atlas & Size & Response & \# Classes  \\\midrule
    HIV & fMRI  & 70 & AAL 116 & $90 \times 90$ & Disease & 2\\
    PNC & fMRI  & 503 & Power 264 & $232 \times 232$ & Gender & 2\\
    PPMI & DTI  & 754 & Desikan-Killiany & $84 \times 84$ & Disease & 2\\
    ABCD & fMRI  & 7,901 & HCP 360 & $360 \times 360$ & Gender & 2\\
    \bottomrule
    
\end{tabular}

\label{tab:dataset}
\end{table*}


    


\begin{table*}
	\centering
	\caption{Performance report (\%) of different message passing GNNs in the four-modular design space with other two representative baselines on four datasets. We highlight the best performed one in each module based on AUC, since it is not sensitive to the changes in the class distribution, providing a fair evaluation on unbalanced datasets like PPMI.}
	\resizebox{\linewidth}{!}{
	\begin{tabular}{cccccccccccccc}
	\toprule
	\multirow{2.5}{*}{Module} &\multirow{2.5}{*}{Method} & \multicolumn{3}{c}{HIV} & \multicolumn{3}{c}{PNC} & \multicolumn{3}{c}{PPMI} & \multicolumn{3}{c}{ABCD}\\
	\cmidrule(lr){3-5} \cmidrule(lr){6-8} \cmidrule(lr){9-11} \cmidrule(lr){12-14}
	& & Accuracy & F1 & AUC & Accuracy & F1 & AUC & Accuracy & F1 & AUC & Accuracy & F1 & AUC \\
	\midrule
	\multirow{5}{*}{\makecell{Node\\Features}} 
	& \textit{Identity}   & 50.00{\tiny±0.00} & 33.33{\tiny±0.00} & 46.73{\tiny±10.57} & 57.34{\tiny±0.17} & 36.44{\tiny±0.17} & 52.58{\tiny±4.80} & 79.25{\tiny±0.24} & 44.21{\tiny±0.08} & 59.65{\tiny±6.80} & 49.97{\tiny±0.13} & 33.32{\tiny±0.06}  & 50.00{\tiny±0.20} \\
	& \textit{Eigen}   & 65.71{\tiny±2.86} & 65.45{\tiny±2.69} & 65.31{\tiny±2.89} & 51.40{\tiny±3.92} & 48.63{\tiny±5.42} & 50.18{\tiny±7.57} & 74.09{\tiny±2.77} & 47.36{\tiny±4.26} & 49.21{\tiny±1.58} & 50.79{\tiny±0.82} & 50.79{\tiny±0.83} & 51.18{\tiny±1.16} \\
	& \textit{Degree}   & 44.29{\tiny±5.35} & 35.50{\tiny±6.10} & 42.04{\tiny±4.00} & 63.89{\tiny±2.27} & 59.69{\tiny±3.85} & 70.25{\tiny±4.38} & 79.52{\tiny±2.31} & 49.40{\tiny±5.17} & 59.73{\tiny±4.31} & 63.46{\tiny±1.29} & 63.45{\tiny±1.28} & 68.16{\tiny±1.41} \\
	& \textit{Degree profile}   & 50.00{\tiny±0.00} & 33.33{\tiny±0.00} & 50.00{\tiny±0.00} & 51.40{\tiny±7.21} & 33.80{\tiny±3.21} & 50.00{\tiny±0.00} & 77.02{\tiny±1.97} & 49.45{\tiny±3.51} & 58.65{\tiny±2.44} & 49.92{\tiny±0.11} & 33.30{\tiny±0.05} & 50.00{\tiny±0.00} \\
	& \textit{Connection profile}   & 65.71{\tiny±13.85} & 64.11{\tiny±13.99} & \textbf{75.10{\tiny±16.95}} & 69.83{\tiny±4.15} & 66.20{\tiny±4.74} & \textbf{76.69{\tiny±5.04}} & 77.99{\tiny±2.78} & 52.96{\tiny±4.52} & \textbf{65.77{\tiny±4.09}} & 82.42{\tiny±1.93} & 82.30{\tiny±2.08} & \textbf{91.33{\tiny±0.77}} \\
	\midrule
	\multirow{5}{*}{\makecell{Message\\Passing}}
	& \textit{Edge weighted}   & 50.00{\tiny±0.00} & 33.33{\tiny±0.00} & 49.80{\tiny±4.20} & 64.87{\tiny±5.44} & 59.70{\tiny±7.04} & 69.98{\tiny±4.19} & 79.25{\tiny±0.24} & 44.21{\tiny±0.08} & 62.26{\tiny±2.80} & 74.47{\tiny±1.17} & 74.36{\tiny±1.23} & 82.37{\tiny±1.46} \\
	& \textit{Bin concat}   & 50.00{\tiny±0.00} & 33.33{\tiny±0.00} & 49.39{\tiny±9.25} & 54.74{\tiny±5.88} & 36.42{\tiny±3.97} & 61.68{\tiny±3.91} & 79.25{\tiny±0.24} & 44.21{\tiny±0.08} & 52.67{\tiny±7.16} & 53.72{\tiny±4.97} & 43.26{\tiny±12.43} & 61.86{\tiny±5.79} \\
	& \textit{Edge weight concat}   & 51.43{\tiny±2.86} & 44.36{\tiny±6.88} & 48.16{\tiny±10.13} & 63.68{\tiny±3.31} & 60.27{\tiny±5.97} & 67.34{\tiny±3.02} & 79.25{\tiny±0.24} & 44.21{\tiny±0.08} & 59.72{\tiny±4.65} & 64.59{\tiny±1.30} & 64.30{\tiny±1.43} & 70.63{\tiny±1.02} \\
	& \textit{Node edge concat}   & 65.71{\tiny±13.85} & 64.11{\tiny±13.99} & 75.10{\tiny±16.95} & 69.83{\tiny±4.15} & 66.20{\tiny±4.74} & 76.69{\tiny±5.04} & 77.99{\tiny±2.78} & 52.96{\tiny±4.52} & 65.77{\tiny±4.09} & 82.42{\tiny±1.93} & 82.30{\tiny±2.08} & 91.33{\tiny±0.77} \\
	& \textit{Node concat}   & 70.00{\tiny±15.91} & 68.83{\tiny±17.57} & \textbf{77.96{\tiny±8.20}} & 70.63{\tiny±2.35} & 67.12{\tiny±1.81} & \textbf{78.32{\tiny±1.42}} & 78.41{\tiny±1.62} & 54.46{\tiny±3.08} & \textbf{68.34{\tiny±1.89}} & 80.50{\tiny±2.27} & 80.10{\tiny±2.47} & \textbf{91.36{\tiny±0.92}} \\
	\midrule
	\multirow{5}{*}{\makecell{Message\\Passing\\ w/ Attention}}
	& \textit{Attention weighted}   & 50.00{\tiny±0.00} & 33.33{\tiny±0.00} & 49.80{\tiny±8.52} & 65.09{\tiny±2.21} & 60.74{\tiny±4.89} & 69.79{\tiny±4.24} & 79.25{\tiny±0.24} & 44.21{\tiny±0.08} & 63.24{\tiny±3.77} & 77.74{\tiny±0.97} & 77.70{\tiny±1.01} & 85.10{\tiny±1.10} \\
	& \textit{Edge weighted w/ attn} & 50.00{\tiny±0.00} & 33.33{\tiny±0.00} & 42.04{\tiny±15.63} & 62.90{\tiny±1.22} & 61.14{\tiny±0.57} & 69.74{\tiny±2.37} & 79.25{\tiny±0.24} & 44.21{\tiny±0.08} & 54.92{\tiny±4.80} & 78.04{\tiny±1.96} & 77.81{\tiny±2.33} & 86.86{\tiny±0.63} \\
	& \textit{Attention edge sum}  & 51.43{\tiny±7.00} & 49.13{\tiny±5.65} & 54.49{\tiny±15.67} & 61.51{\tiny±2.86} & 55.36{\tiny±4.76} & 69.38{\tiny±3.50} & 79.11{\tiny±0.40} & 44.17{\tiny±0.12} & 60.47{\tiny±6.26} & 75.71{\tiny±1.52} & 75.59{\tiny±1.68} & 83.78{\tiny±0.82} \\
	& \textit{Node edge concat w/ attn}  & 72.86{\tiny±11.43} & 72.52{\tiny±11.72} & 78.37{\tiny±10.85} & 67.66{\tiny±5.07} & 64.69{\tiny±5.36} & 74.52{\tiny±1.20} & 77.30{\tiny±1.52} & 50.96{\tiny±4.20} & 63.93{\tiny±4.89} & 83.10{\tiny±0.47} & 83.03{\tiny±0.52} & \textbf{91.85{\tiny±0.29}} \\
	& \textit{Node concat  w/ attn} & 71.43{\tiny±9.04} & 70.47{\tiny±9.26} & \textbf{82.04{\tiny±11.21}} & 68.85{\tiny±6.42} & 64.29{\tiny±10.15} & \textbf{75.36{\tiny±5.09}} & 78.41{\tiny±1.43} & 49.98{\tiny±1.87} & \textbf{68.14{\tiny±5.01}} & 83.19{\tiny±0.93} & 83.12{\tiny±0.96} & 91.55{\tiny±0.59} \\
	\midrule 
	\multirow{4}{*}{\makecell{Pooling\\Strategies}}
	& \textit{Mean pooling}   & 47.14{\tiny±15.39} & 41.71{\tiny±17.36} & 58.78{\tiny±18.63} & 66.86{\tiny±2.33} & 61.39{\tiny±4.88} & 74.20{\tiny±3.39} & 79.25{\tiny±0.24} & 44.21{\tiny±0.08} & 59.64{\tiny±5.47} & 81.13{\tiny±0.35} & 81.06{\tiny±0.34} & 88.49{\tiny±1.12} \\
	& \textit{Sum pooling}   & 57.14{\tiny±9.04} & 52.23{\tiny±12.65} & 57.96{\tiny±11.15} & 60.13{\tiny±2.87} & 53.96{\tiny±7.61} & 66.11{\tiny±4.22} & 79.39{\tiny±0.52} & 47.68{\tiny±3.12} & 61.29{\tiny±2.11} & 77.48{\tiny±3.75} & 76.96{\tiny±4.58} & 87.90{\tiny±0.65} \\
	& \textit{Concat pooling}   & 65.71{\tiny±13.85} & 64.11{\tiny±13.99} & 75.10{\tiny±16.95} & 69.83{\tiny±4.15} & 66.20{\tiny±4.74} & \textbf{76.69{\tiny±5.04}} & 77.99{\tiny±2.78} & 52.96{\tiny±4.52} & \textbf{65.77{\tiny±4.09}} & 82.42{\tiny±1.93} & 82.30{\tiny±2.08} & \textbf{91.33{\tiny±0.77}} \\
	& \textit{\textcolor{black}{DiffPool}} & \textcolor{black}{72.86{\tiny±21.19}} & \textcolor{black}{70.22{\tiny±23.91}} & \textcolor{black}{\textbf{76.57{\tiny±17.16}}} & \textcolor{black}{62.72{\tiny±12.40}} & \textcolor{black}{75.95{\tiny±4.28}} & \textcolor{black}{64.08{\tiny±16.71}}  & \textcolor{black}{78.42{\tiny±3.53}} & \textcolor{black}{56.55{\tiny±6.48}} & \textcolor{black}{63.07{\tiny±7.77}} & \textcolor{black}{76.45{\tiny±1.44}} & \textcolor{black}{76.35{\tiny±1.52}} & \textcolor{black}{83.92{\tiny±1.25}} \\
		
	\cmidrule{1-14}\morecmidrules\cmidrule{1-14}
	\multirow{3}{*}{\makecell{Shallow\\Baselines}}
	& \textcolor{black}{M2E} & \textcolor{black}{57.14{\tiny±19.17}} & \textcolor{black}{53.71{\tiny±19.80}} & \textcolor{black}{57.50{\tiny±18.71}} & \textcolor{black}{53.76{\tiny±4.94}} & \textcolor{black}{46.10{\tiny±6.94}} & \textcolor{black}{49.70{\tiny±5.18}}  & \textcolor{black}{78.69{\tiny±1.78}} & \textcolor{black}{45.81{\tiny±4.17}} & \textcolor{black}{50.39{\tiny±2.59}} & \textcolor{black}{50.10{\tiny±1.90}} & \textcolor{black}{49.95{\tiny±1.88}} & \textcolor{black}{50.10{\tiny±1.90}} \\
	& \textcolor{black}{MPCA}   & \textcolor{black}{67.14{\tiny±20.25}} & \textcolor{black}{64.28{\tiny±23.47}} & \textcolor{black}{69.17{\tiny±20.17}} & \textcolor{black}{76.76{\tiny±4.30}} & \textcolor{black}{75.95{\tiny±4.28}} & \textcolor{black}{76.05{\tiny±4.34}}  & \textcolor{black}{79.15{\tiny±0.57}} & \textcolor{black}{44.18{\tiny±0.18}} & \textcolor{black}{50.00{\tiny±0.00}} & \textcolor{black}{88.94{\tiny±1.64}} & \textcolor{black}{88.94{\tiny±1.64}} & \textcolor{black}{88.94{\tiny±1.64}} \\
	& \textcolor{black}{MK-SVM}   & \textcolor{black}{65.71{\tiny±7.00}} & \textcolor{black}{62.08{\tiny±7.49}} & \textcolor{black}{65.83{\tiny±7.41}} & \textcolor{black}{78.38{\tiny±5.09}} & \textcolor{black}{77.55{\tiny±5.83}} & \textcolor{black}{77.57{\tiny±5.65}}  & \textcolor{black}{79.15{\tiny±0.57}} & \textcolor{black}{44.18{\tiny±0.18}} & \textcolor{black}{50.00{\tiny±0.00}} & \textcolor{black}{89.42{\tiny±0.97}} & \textcolor{black}{89.42{\tiny±0.97}} & \textcolor{black}{89.42{\tiny±0.97}} \\
	\midrule
	\multirow{2}{*}{\makecell{Deep\\Baselines}}
	& BrainNetCNN   & 60.21{\tiny±17.16} & 60.12{\tiny±13.56} & 70.93{\tiny±4.01} & 71.93{\tiny±4.90}  & 69.94{\tiny±5.42} & 78.50{\tiny±3.28} & 77.24{\tiny±2.09} & 50.24{\tiny±3.09} & {58.76\tiny±8.95} & {85.1{\tiny±0.92}} & {85.7{\tiny±0.83}} & {93.5{\tiny±0.34}} \\
	& BrainGNN   & 62.98{\tiny±11.15} & 60.45{\tiny±8.96} & 68.03{\tiny±9.16} & 70.62{\tiny±4.85} & 68.93{\tiny±4.01} & 77.53{\tiny±3.23}  & 79.17{\tiny±1.22} & 44.19{\tiny±3.11} & 45.26{\tiny±3.65} & {OOM} & {OOM} & {OOM} \\
	\bottomrule
	\end{tabular}
	}
	\label{tab:performance}
\end{table*}

\label{sec:exp}
In this section, we show experimental results on brain networks generated from real-world neuroimaging studies with different GNN modular designs. Varying each design dimension under each module results in a total of 375 different architectures. Note that here we do not aim to cover all combinations, but to quickly find a relatively good one. Furthermore, we emphasize that the design space can be expanded as new design dimensions emerge. 

\subsection{Experimental Settings}
\subsubsection{Datasets}
\hejie{To establish a benchmark for generic brain network analysis models, we include four datasets processed and constructed from different neuroimaging modalities, specifically fMRI (HIV \cite{liu2018multi}, PNC\footnote{\url{https://www.nitrc.org/projects/pnc}}, ABCD\footref{foot:abcd}) and dMRI (PPMI\footref{foot:ppmi}), based on different brain atlas. For the HIV and PPMI datasets, the task is to classify patients from healthy control (Patient, Normal Control); while for the PNC and ABCD datasets, the task is gender prediction (Male, Female). We intentionally cover such a diverse set of datasets from different modalities (and preprocessing procedures/parcellations/tasks), because our purpose is to establish a benchmark for generic brain network analysis models. Thus observations on a diverse set of datasets can be more instructive for methodology focused studies. All the datasets we used have been visually checked by imaging experts in our team for quality control. Among these four datasets, PNC, PPMI, and ABCD are restrictively publicly available ones that can be requested and downloaded from their official website. The dataset information is summarized in TABLE \ref{tab:dataset}. Since the datasets can be acquired from multiple sites, multisite issues need to be addressed when performing the analysis on the constructed networks. Over the past few years, ComBat techniques \cite{chen2020removal, fortin2018harmonization} from the microarray literature have started to be used more frequently to deal with multi-site batch effects. Since our benchmark focuses more on a comprehensive overview of brain network construction and effective GNN designs for brain networks, advanced methods for handling multi-site issues are out of the scope of this work. Interested readers can refer to \cite{chen2014exploration, bell2022harmonization, pomponio2020harmonization, yamashita2019harmonization, pinto2020harmonization} for more advanced multisite data handling methods. } 

\begin{itemize}
    \item \textit{Human Immunodeficiency Virus Infection (HIV): } This dataset is collected from the Chicago Early HIV Infection Study at Northwestern University. The clinical cohort includes fMRI imaging of 70 subjects, 35 of which are early HIV patients and the other 35 are seronegative controls. The preprocessing includes realignment to the first volume, followed by slice timing correction, normalization, and spatial smoothness, band-pass filtering, and linear trend removal of the time series. We focus on the 116 anatomical ROIs \cite{tzourio2002automated} and extract a sequence of time courses from them. Finally, brain networks with 90 cerebral regions are constructed, with links representing the correlations between ROIs.
    \item \textit{Philadelphia Neuroimaging Cohort (PNC): } This rs-fMRI dataset is from the Brain Behavior Laboratory at the University of Pennsylvania and the Children's Hospital of Philadelphia. 289 (57.46\%) of the 503 included subjects are female, indicating this dataset is balanced across genders. The regions are parcellated based on the 264-node atlas defined by \citet{power264}. The preprocessing includes slice timing correction, motion correction, registration, normalization, removal of linear trends, bandpass filtering, and spatial smoothing. In the resulting data, each sample contains 264 nodes with time-series data collected through 120 time steps. We focus on the 232 nodes in the Power's atlas associated with major resting-state functional modules \citep{smith2009correspondence}.
    \item \textit{Parkinson's Progression Markers Initiative (PPMI): } This dataset is from a collaborative study for Parkinson's Research to improve PD therapeutics. We consider the DTI acquisition of 754 subjects, with 596 Parkinson's disease patients and 158 healthy controls. The raw data are first aligned to correct for head motion and eddy current distortions. Then the non-brain tissue is removed and the skull-stripped images are linearly aligned and registered. 84 ROIs are parcellated from T1-weighted structural MRI \hejie{based on the Desikan-Killiany' cortical atlas \cite{desikan2006automated}} and the brain network is reconstructed using the deterministic 2nd-order Runge-Kutta (RK2) whole-brain tractography algorithm \cite{zhan2015comparison}.
    \item \textit{Adolescent Brain Cognitive Development Study (ABCD)}: This study recruits children aged 9-10 years across 21 sites in the U.S. Each child is followed into early adulthood, with repeated imaging scans, as well as extensive psychological and cognitive tests \cite{casey2018adolescent}. After selection, 7,901 children are included in the analysis, with 3,961 (50.1\%) female. We use rs-fMRI scans for the baseline visit processed with the standard and open-source ABCD-HCP BIDS fMRI Pipeline\footref{foot:abcdpip}. After processing, each sample contains a connectivity matrix whose size is $360 \times 360$ and BOLD time-series for each node. The region definition is based on the HCP 360 ROI atlas \citep{GLASSER2013105}.
\end{itemize}

\hejie{Structural connectivity and functional connectivity are different in their strength and sparsity, thus need to be handled differently. For structural connectivity, we normalize the edge weights by dividing each value by the maximum value in a sample. The processed edge weights are thus ranged from 0 to 1. For functional connectivity, we follow common practice to remove the negative values for GNNs that cannot handle negative values (like GCN), and keep them for GNNs that can handle negative values (like GAT).}

\subsubsection{Baselines}
\hejie{For comprehensiveness, we compare our modular design with competitors of both shallow and deep models. The shallow methods we consider include M2E \cite{Liu:2018ty}, MPCA \cite{Lu:2008cw}, and MK-SVM \cite{Dyrba:2015ci}, where the output graph-level embeddings are evaluated using logistic regression classifiers. Specifically, M2E is a partially-symmetric tensor factorization based method for brain network analysis, and it has been empirically compared with spectral embedding clustering methods such as SEC \cite{nie2011spectral} or spectral learning frameworks such as AMGL \cite{nie2016parameter}; MPCA is proposed for the feature extraction and analysis of tensor objects such as neuroimaging; multiple kernel SVM (MK-SVM) is essentially an extension of the conventional SVM algorithm and has been applied for the analysis of functional and structural connectivity in Alzheimer’s disease. We also include two state-of-the-art deep models specifically designed for brain networks: BrainGNN \cite{li2020braingnn} and BrainNetCNN \cite{kawahara2017brainnetcnn}. \minor{The message passing in BrainGNN is Edge weighted and it further leverages additional regional information (such as coordinates or ROI ordering based one-hot embeddings) to assign a separate GCN kernel for each ROI where ROIs in the same community are embedded by the similar kernel and those in different communities are embedded in different ways, but this will introduce a lot of additional model parameters and make the model hard to train}. On the other hand, BrainNetCNN models the adjacency matrix of a brain network as a 2D image and does not follow the message passing mechanism as we discussed in Section \ref{mp}. Note that the purpose of our paper, and of most benchmark papers, is not to establish superior performance of a certain method, but rather to provide an effective and fair ground for comparing different methods.}
\subsubsection{Implementation Details}
The proposed model is implemented using PyTorch 1.10.2 \cite{Paszke:2019vf} and PyTorch Geometric 2.0.3 \cite{Fey:2019wv}. A Quadro RTX 8000 GPU with 48GB of memory is used for model training. \hejie{The optimizer we used is Adam. We train all of our models through 20 epochs, and the learning rate is 1e-3. We use a weight decay of 1e-4 as a means of regularization. The loss function is cross entropy.} Hyper-parameters are selected automatically with an open-source AutoML toolkit NNI\footnote{\url{https://github.com/microsoft/nni}\label{nni}}. Please refer to our repository for comprehensive parameter configurations. \hejie{When tuning the hyperparameters, we first split the dataset into a train set and a test set with the ratio of 8:2. The k-fold validation is performed on the train set, where we further divide the train set into 10 parts and take one in each run to use as the validation set. The selection of the best hyperparameter is based on the average performance of the model on the validation sets. The reported metrics in Table II, on the other hand, is the average performance on the test set, with each run trained on different train sets. The competing methods are also tuned in the same way. For BrainGNN, we used the author's open-source code\footnote{\url{https://github.com/xxlya/BrainGNN_Pytorch}\label{braingnn}}. For BrainNetCNN, we implemented it by ourselves with PyTorch, which is publicly available in our BrainGB package\footnote{\url{https://github.com/HennyJie/BrainGB}}. For the hyper-parameter tuning, we selected several important hyper-parameters and performed the grid search on them based on the provided best setting as claimed in their paper. To be specific, for BrainGNN, we searched for different learning rates in \{0.01, 0.005, 0.001\} with different feature dimensions in \{100, 200\} and the number of GNN layers in \{2, 3\}. For BrainNetCNN, we searched for different dropout rates in \{0.3, 0.5, 0.7\} with learning rates in \{0.001, 0.0005, 0.0001\} and the number of layers in MLP in \{1, 2, 3\}. The reported results of these two baselines in Table II are from the best performing groups, where for BrainGNN, the learning rate is 0.01, the feature dimension is 200 and the number of GNN layers is 2, and for BrainNetCNN, the dropout rate is 0.3, the learning rate is 0.0001 and the number of layers in MLP is 3.} The metrics used to evaluate performance are Accuracy, F1 score, and Area Under the ROC Curve (AUC), which are widely used for disease identification. To indicate the robustness of each model, all the reported results are the average performance of ten-fold cross-validation conducted on different train/test splits. 
 
\subsection{Performance Report} 
\subsubsection{Node Feature}
On comparing node features, we set the other modules as the well-performed settings in individual tests. Specifically, we use \textit{node edge concat} in Eq. \ref{eq:edge_node_concat} as the message passing scheme, and \textit{concat pooling} in Eq. \ref{concat} as the pooling strategy. 
Our experimental results demonstrate that the \textbf{\textit{connection profile}} which uses the corresponding row in the adjacency matrix as the node features achieves the best performance across all datasets, with up to 33.99\% improvements over the second-best, \textit{degree}, on ABCD. We believe this is because the \textit{connection profile} captures the whole picture of structural information in the brain network, and preserves rich information on pairwise connections that can be used to perform brain parcellation.
In general, the structure node features (\eg, \textit{degree}, \textit{connection profile}) perform  better than the positional ones (\eg, \textit{identity}, \textit{eigen}), indicating that the overall structural information of graph and the structural role of each node are important in the task of brain network analysis. This conclusion is consistent with previous findings in the literature that structural artificial node features work well for graph-level tasks on general graphs \cite{cui2021positional}.

\subsubsection{Message Passing}
To study the effectiveness of different message passing schemes, we initialize the node features with \textit{connection profile} and apply the \textit{concat pooling} to produce graph-level representations, which both perform best when examined separately in each module. 
Our results reveal that \textbf{\textit{node concat}} (Eq. \ref{eq:node_concat}) message passing has the highest AUC performance across four datasets, followed by \textit{node edge concat} (Eq. \ref{eq:edge_node_concat}), which achieves a similar AUC performance with sometimes slightly better accuracy and F1 scores (ABCD). The performance superiority of the last two methods may arise from their advantage of reinforcing self-representation of the central node during each step of message passing. This helps to retain the original information from the last step and avoid over-fitting towards a biased direction in the optimization process. Surprisingly, the edge involved \textit{node edge concat} performs slightly worse than the pure \textit{node concat}, though the gap gets closer on larger datasets. This indicates that encoding edge weights as a single value may not be useful when the global structure has already been used as the initial node features.

\subsubsection{Attention Enhanced Message Passing}
When evaluating the effectiveness of different attention-enhanced message passing schemes, we set the node features as \textit{connection profile} and apply the \textit{concat pooling} strategy, just as for the evaluation of message passing without attention mechanisms. It is shown that the \textbf{\textit{node concat w/ attn}} (Eq. \ref{Node concat w/ attn}) and \textbf{\textit{node edge concat w/ attn}} (Eq. \ref{Node edge concat w/ attn}) yield very close results across four datasets and they alternately perform the best. Furthermore, the attention-enhanced version achieves better outcomes most of the time (up to 5.23\% relative improvements) vs. the corresponding message passing architecture without an attention mechanism. This demonstrates the effectiveness of utilizing learnable attention weights in the GNN aggregation and update process in addition to the fixed edge weights. Also, the \textit{node edge concat w/ attn} surpasses \textit{node concat w/ attn} on the larger dataset (e.g., ABCD), which may imply potential advantages of involving edge weights into message design when there are enough training samples. 

\subsubsection{Pooling Strategies}
For studying pooling strategies, we employ the \textit{node edge concat} (Eq. \ref{eq:edge_node_concat}) as the message passing scheme and \textit{connection profile} as the initial node features. Our findings reveal that the \textbf{\textit{concat pooling}} strategy (Eq. \ref{concat}) consistently outperforms the other two methods across all four datasets. This is likely because when \textit{concat} is used, the final node representations of all the brain regions are kept in the graph-level representation for classifiers. The other two paradigms, on the other hand, obtain a graph-level embedding with the same dimension of node features. Thus they lose some information that could be helpful for graph-level prediction tasks. Though \textit{concat} does not ensure permutation invariance, it is actually not needed for brain network analysis since the node order given a parcellation is fixed. \hejie{The compared hierarchical pooling method \textit{DiffPool} demonstrates some advantages on the small HIV dataset but fails to surpass the simple \textit{concat} pooling on three other larger datasets.}


\subsubsection{Other Baselines}
\hejie{In general, we expect deep models like GNNs to perform better on larger datasets. For example, the performance of GNN models on the ABCD dataset clearly surpasses all shallow models by about 2 percent. However, this trend should not prohibit one from experimenting with GNN models on smaller datasets. GNNs do perform well on some small datasets, such as the HIV dataset. Despite running on a small dataset, GNN models in BrainGB have an over 5 percent advantage over all shallow models.} \hejie{As for the deep baselines, BrainGNN can be out-of-memory (OOM) on large datasets. The best combination based on our modular design outperforms BrainGNN on all four datasets (HIV, PNC, PPMI and ABCD) and achieves comparable results with BrainNetCNN in most cases especially on smaller datasets (HIV, PPMI).} These findings prove the need to carefully experiment with our modular designs of GNNs before further developing more complicated architectures, which might just overfit certain datasets.

\subsubsection{Insights on Density Levels}
\hejie{Functional connectivity and structural connectivity have distinctive differences in sparsity levels. Functional networks like ABCD are fully connected. Structural networks like PPMI contain approximately 22.64\% edges on average. Through our experiments, we found sparsity levels do have an impact on the choices of hyperparameters. For example, GNNs on the sparser structural networks of PPMI reach the maximum performance with a hidden dimension of 64, whereas on the functional network of ABCD, they have an optimal hidden dimension of 256, which indicates that GNN models should more complicated with more learnable parameters when the input networks are denser. This observation can be instructive for designing GNN architectures on brain networks constructed from different modalities. }
\section{Open Source Benchmark Platform}
\label{sec:open}
To foster future research, we provide an out-of-box package that can be directly installed through pip, with installation and tutorials on our hosted \benchmark website \url{https://braingb.us}. The \benchmark package is also open-sourced at \url{https://github.com/HennyJie/BrainGB}. We provide examples of GNN-based brain network analysis, trained models, and instructions on imaging preprocessing and functional and structural brain networks construction from raw fMRI and dMRI respectively. 
It is noted that due to the modular designs, \benchmark can also be extended to other tasks, by adding task-specific functions in each module. 
\section{Discussion and Extensions}
\label{sec:conclu}

In this paper, we first present \benchmark, a \textit{unified}, \textit{modular}, \textit{scalable}, and \textit{reproducible} framework for brain network analysis with GNNs.
While the dataset generation, baselines, and evaluations we provide in BrainGB are thorough, we consider several limitations in the current paradigm:
\begin{itemize}
\item The aggregation mechanism in GNN is known to be effective for node-level tasks with the effect of node feature smoothing, and for graph-level tasks due to its capability in structure differentiation. However, for brain networks, what kinds of graph structures (\eg, communities, subgraphs) are effective beyond the pairwise connections are still unknown. 
\item The small size of neuroimaging datasets may limit the effectiveness and generalization ability of complex deep learning models. 
\end{itemize}
Towards these two limitations, we envision several future directions that can be potentially helpful to fully unleash the power of GNNs for brain network analysis: 
\begin{itemize}
\item Neurology-driven GNN designs: to design the GNN architectures based on neurological understandings of predictive brain signals, especially disease-specific ones. 
\item Pre-training and transfer learning of GNNs: to design techniques that can train complex GNN models across studies and cohorts \cite{yang2022data}. Besides, information sharing across different diseases could lead to a better understanding of cross-disorder commonalities.
\end{itemize}
\section*{Acknowledgment}

This research was supported in part by the University Research Committee of Emory University, and the internal funding and GPU servers provided by the Computer Science Department of Emory University. The authors gratefully acknowledge support from National Institutes of Health (R01MH105561, R01MH118771, R01AG071243, R01MH125928, U01AG068057), National Science Foundation (IIS 2045848, IIS 1837956) and Office of Naval Research (N00014-18-1-2009). The content is solely the responsibility of the authors and does not necessarily represent the official views of the NIH, NSF, and ONR. 

Support for the collection of the Philadelphia Neurodevelopmental Cohort (PNC) dataset was provided by grant RC2MH089983 awarded to Raquel Gur and RC2MH089924 awarded to Hakon Hakorson. 
The ABCD Study\textsuperscript{\textregistered} is supported by the National Institutes of Health and additional federal partners under award numbers \seqsplit{U01DA041048,~U01DA050989,~U01DA051016,~U01DA041022,~U01DA051018,~U01DA051037,~U01DA050987,~U01DA041174,~U01DA041106,~U01DA041117,~U01DA041028,~U01DA041134,~U01DA050988,~U01DA051039,~U01DA041156,~U01DA041025,~U01DA041120,~U01DA051038,~U01DA041148,~U01DA041093,~U01DA041089,~U24DA041123,~U24DA041147}. A full list of supporters is available at  \sloppy\url{https://abcdstudy.org/federal-partners.html}. A listing of participating sites and a complete listing of the study investigators can be found at  \url{https://abcdstudy.org/consortium_members/}. This manuscript reflects the views of the authors and may not reflect the opinions or views of the NIH or ABCD consortium investigators. The ABCD data repository grows and changes over time. The ABCD data used in this report came from NIMH Data Archive Release 4.0 (DOI 10.15154/1523041). DOIs can be found at \url{https://nda.nih.gov/abcd}.

\bibliographystyle{IEEEtranN}
\balance
\bibliography{reference.bib}

\begin{thebibliography}{107}
\providecommand{\natexlab}[1]{#1}
\providecommand{\url}[1]{#1}
\csname url@samestyle\endcsname
\providecommand{\newblock}{\relax}
\providecommand{\bibinfo}[2]{#2}
\providecommand{\BIBentrySTDinterwordspacing}{\spaceskip=0pt\relax}
\providecommand{\BIBentryALTinterwordstretchfactor}{4}
\providecommand{\BIBentryALTinterwordspacing}{\spaceskip=\fontdimen2\font plus
\BIBentryALTinterwordstretchfactor\fontdimen3\font minus
  \fontdimen4\font\relax}
\providecommand{\BIBforeignlanguage}[2]{{%
\expandafter\ifx\csname l@#1\endcsname\relax
\typeout{** WARNING: IEEEtranN.bst: No hyphenation pattern has been}%
\typeout{** loaded for the language `#1'. Using the pattern for}%
\typeout{** the default language instead.}%
\else
\language=\csname l@#1\endcsname
\fi
#2}}
\providecommand{\BIBdecl}{\relax}
\BIBdecl

\bibitem[Li et~al.(2021)Li, Zhou, Dvornek, Zhang, Gao, Zhuang, Scheinost,
  Staib, Ventola, and Duncan]{li2020braingnn}
X.~Li, Y.~Zhou, N.~Dvornek, M.~Zhang, S.~Gao, J.~Zhuang, D.~Scheinost, L.~H.
  Staib, P.~Ventola, and J.~S. Duncan, ``Braingnn: Interpretable brain graph
  neural network for fmri analysis,'' \emph{Med Image Anal}, 2021.

\bibitem[Farahani et~al.(2019)Farahani, Karwowski, and
  Lighthall]{farahani2019application}
F.~V. Farahani, W.~Karwowski, and N.~R. Lighthall, ``Application of graph
  theory for identifying connectivity patterns in human brain networks: a
  systematic review,'' \emph{Front. Neurosci.}, vol.~13, p. 585, 2019.

\bibitem[Osipowicz et~al.(2016)Osipowicz, Sperling, Sharan, and
  Tracy]{osipowicz2016functional}
K.~Osipowicz, M.~R. Sperling, A.~D. Sharan, and J.~I. Tracy, ``Functional mri,
  resting state fmri, and dti for predicting verbal fluency outcome following
  resective surgery for temporal lobe epilepsy,'' \emph{J. Neurosurg.}, vol.
  124, pp. 929--937, 2016.

\bibitem[Maglanoc et~al.(2020)Maglanoc, Kaufmann, Jonassen, Hilland, Beck,
  Landr{\o}, and Westlye]{maglanoc2020multimodal}
L.~A. Maglanoc, T.~Kaufmann, R.~Jonassen, E.~Hilland, D.~Beck, N.~I. Landr{\o},
  and L.~T. Westlye, ``Multimodal fusion of structural and functional brain
  imaging in depression using linked independent component analysis,''
  \emph{Hum Brain Mapp}, vol.~41, pp. 241--255, 2020.

\bibitem[Bullmore and Sporns(2009)]{bullmore2009complex}
E.~Bullmore and O.~Sporns, ``Complex brain networks: graph theoretical analysis
  of structural and functional systems,'' \emph{Nat. Rev. Neurosci.}, vol.~10,
  pp. 186--198, 2009.

\bibitem[Sporns(2022)]{sporns2022graph}
O.~Sporns, ``Graph theory methods: applications in brain networks,''
  \emph{Dialogues Clin. Neurosci.}, 2022.

\bibitem[Liu et~al.(2018{\natexlab{a}})]{Liu:2018ty}
Y.~Liu \emph{et~al.}, ``Multi-view multi-graph embedding for brain network
  clustering analysis,'' in \emph{AAAI}, 2018.

\bibitem[Zhan et~al.(2015{\natexlab{a}})Zhan, Liu, Wang, Zhou, Jahanshad, Ye,
  Thompson, and (ADNI)]{zhan2015boosting}
L.~Zhan, Y.~Liu, Y.~Wang, J.~Zhou, N.~Jahanshad, J.~Ye, P.~M. Thompson, and
  A.~D. N.~I. (ADNI), ``Boosting brain connectome classification accuracy in
  alzheimer's disease using higher-order singular value decomposition,''
  \emph{Frontiers in neuroscience}, vol.~9, p. 257, 2015.

\bibitem[Faskowitz et~al.(2021)Faskowitz, Betzel, and
  Sporns]{faskowitz2021edges}
J.~Faskowitz, R.~F. Betzel, and O.~Sporns, ``Edges in brain networks:
  Contributions to models of structure and function,'' \emph{arXiv.org}, 2021.

\bibitem[Dosovitskiy et~al.(2021)Dosovitskiy, Beyer, Kolesnikov, Weissenborn,
  Zhai, Unterthiner, Dehghani, Minderer, Heigold, Gelly, Uszkoreit, and
  Houlsby]{DBLP:conf/iclr/DosovitskiyB0WZ21}
A.~Dosovitskiy, L.~Beyer, A.~Kolesnikov, D.~Weissenborn, X.~Zhai,
  T.~Unterthiner, M.~Dehghani, M.~Minderer, G.~Heigold, S.~Gelly, J.~Uszkoreit,
  and N.~Houlsby, ``An image is worth 16x16 words: Transformers for image
  recognition at scale,'' in \emph{ICLR}, 2021.

\bibitem[Radford et~al.(2021)Radford, Kim, Hallacy, Ramesh, Goh, Agarwal,
  Sastry, Askell, Mishkin, Clark, et~al.]{radford2021learning}
A.~Radford, J.~W. Kim, C.~Hallacy, A.~Ramesh, G.~Goh, S.~Agarwal, G.~Sastry,
  A.~Askell, P.~Mishkin, J.~Clark \emph{et~al.}, ``Learning transferable visual
  models from natural language supervision,'' in \emph{ICML}, 2021.

\bibitem[Arnab et~al.(2021)Arnab, Dehghani, Heigold, Sun, Lu{\v{c}}i{\'c}, and
  Schmid]{arnab2021vivit}
A.~Arnab, M.~Dehghani, G.~Heigold, C.~Sun, M.~Lu{\v{c}}i{\'c}, and C.~Schmid,
  ``Vivit: A video vision transformer,'' in \emph{ICCV}, 2021.

\bibitem[Gulati et~al.(2020)Gulati, Qin, Chiu, Parmar, Zhang, Yu, Han, Wang,
  Zhang, Wu, and Pang]{DBLP:conf/interspeech/GulatiQCPZYHWZW20}
A.~Gulati, J.~Qin, C.~Chiu, N.~Parmar, Y.~Zhang, J.~Yu, W.~Han, S.~Wang,
  Z.~Zhang, Y.~Wu, and R.~Pang, ``Conformer: Convolution-augmented transformer
  for speech recognition,'' in \emph{INTERSPEECH}, 2020.

\bibitem[Kipf and Welling(2017)]{kipf2016semi}
T.~N. Kipf and M.~Welling, ``Semi-supervised classification with graph
  convolutional networks,'' in \emph{ICLR}, 2017.

\bibitem[Xu et~al.(2019)Xu, Hu, Leskovec, and Jegelka]{xu2019powerful}
K.~Xu, W.~Hu, J.~Leskovec, and S.~Jegelka, ``How powerful are graph neural
  networks?'' in \emph{ICLR}, 2019.

\bibitem[Veli{\v{c}}kovi{\'c} et~al.(2018)Veli{\v{c}}kovi{\'c}, Cucurull,
  Casanova, Romero, Lio, and Bengio]{velivckovic2018graph}
P.~Veli{\v{c}}kovi{\'c}, G.~Cucurull, A.~Casanova, A.~Romero, P.~Lio, and
  Y.~Bengio, ``Graph attention networks,'' in \emph{ICLR}, 2018.

\bibitem[Kawahara et~al.(2017)Kawahara, Brown, Miller, Booth, Chau, Grunau,
  Zwicker, and Hamarneh]{kawahara2017brainnetcnn}
J.~Kawahara, C.~J. Brown, S.~P. Miller, B.~G. Booth, V.~Chau, R.~E. Grunau,
  J.~G. Zwicker, and G.~Hamarneh, ``Brainnetcnn: Convolutional neural networks
  for brain networks; towards predicting neurodevelopment,'' \emph{NeuroImage},
  vol. 146, pp. 1038--1049, 2017.

\bibitem[Murugesan et~al.(2020)Murugesan, Ganesh, Nalawade, Davenport, Wagner,
  Kim, and Maldjian]{murugesan2020brainnet}
G.~K. Murugesan, C.~Ganesh, S.~Nalawade, E.~M. Davenport, B.~Wagner, W.~H. Kim,
  and J.~A. Maldjian, ``Brainnet: Inference of brain network topology using
  machine learning,'' \emph{Brain Connect}, vol.~10, pp. 422--435, 2020.

\bibitem[Su et~al.(2020)Su, Xu, Pathak, and Wang]{su2020deep}
C.~Su, Z.~Xu, J.~Pathak, and F.~Wang, ``Deep learning in mental health outcome
  research: a scoping review,'' \emph{Transl. Psychiatry}, vol.~10, pp. 1--26,
  2020.

\bibitem[Satterthwaite et~al.(2015)Satterthwaite, Wolf, Roalf, Ruparel, Erus,
  Vandekar, Gennatas, Elliott, Smith, Hakonarson,
  et~al.]{satterthwaite2015linked}
T.~D. Satterthwaite, D.~H. Wolf, D.~R. Roalf, K.~Ruparel, G.~Erus, S.~Vandekar,
  E.~D. Gennatas, M.~A. Elliott, A.~Smith, H.~Hakonarson \emph{et~al.},
  ``Linked sex differences in cognition and functional connectivity in youth,''
  \emph{Cereb. Cortex}, vol.~25, pp. 2383--2394, 2015.

\bibitem[Deco et~al.(2011)Deco, Jirsa, and McIntosh]{deco2011emerging}
G.~Deco, V.~K. Jirsa, and A.~R. McIntosh, ``Emerging concepts for the dynamical
  organization of resting-state activity in the brain,'' \emph{Nat. Rev.
  Neurosci.}, vol.~12, pp. 43--56, 2011.

\bibitem[Wang and Guo(2019)]{wang2019hierarchical}
Y.~Wang and Y.~Guo, ``A hierarchical independent component analysis model for
  longitudinal neuroimaging studies,'' \emph{NeuroImage}, vol. 189, pp.
  380--400, 2019.

\bibitem[Yu et~al.(2019)Yu, Qiao, Chen, Lee, Fei, and Shen]{yu2019weighted}
R.~Yu, L.~Qiao, M.~Chen, S.-W. Lee, X.~Fei, and D.~Shen, ``Weighted graph
  regularized sparse brain network construction for mci identification,''
  \emph{Pattern Recognit}, vol.~90, pp. 220--231, 2019.

\bibitem[Insel and Cuthbert(2015)]{insel2015brain}
T.~R. Insel and B.~N. Cuthbert, ``Brain disorders? precisely,'' \emph{Science},
  vol. 348, pp. 499--500, 2015.

\bibitem[Williams(2016)]{williams2016precision}
L.~M. Williams, ``Precision psychiatry: a neural circuit taxonomy for
  depression and anxiety,'' \emph{Lancet Psychiatry}, vol.~3, pp. 472--480,
  2016.

\bibitem[Li et~al.(2016)Li, Wang, Li, Huang, and Chen]{li2016novel}
W.~Li, M.~Wang, Y.~Li, Y.~Huang, and X.~Chen, ``A novel brain network
  construction method for exploring age-related functional reorganization,''
  \emph{Comput. Intell. Neurosci.}, vol. 2016, 2016.

\bibitem[Zimmermann et~al.(2018)Zimmermann, Griffiths, and
  McIntosh]{zimmermann2018unique}
J.~Zimmermann, J.~D. Griffiths, and A.~R. McIntosh, ``Unique mapping of
  structural and functional connectivity on cognition,'' \emph{J. Neurosci.},
  vol.~38, pp. 9658--9667, 2018.

\bibitem[Hu et~al.(2022)Hu, Zeydabadinezhad, Li, and Guo]{hu2022multimodal}
Y.~Hu, M.~Zeydabadinezhad, L.~Li, and Y.~Guo, ``A multimodal multilevel
  neuroimaging model for investigating brain connectome development,'' \emph{J
  Am Stat Assoc}, pp. 1--15, 2022.

\bibitem[Martensson et~al.(2018)Martensson, Pereira, Mecocci, Vellas, Tsolaki,
  K{\l}oszewska, Soininen, Lovestone, Simmons, Volpe,
  et~al.]{maartensson2018stability}
G.~Martensson, J.~B. Pereira, P.~Mecocci, B.~Vellas, M.~Tsolaki,
  I.~K{\l}oszewska, H.~Soininen, S.~Lovestone, A.~Simmons, G.~Volpe
  \emph{et~al.}, ``Stability of graph theoretical measures in structural brain
  networks in alzheimer’s disease,'' \emph{Sci. Rep.}, vol.~8, pp. 1--15,
  2018.

\bibitem[Yahata et~al.(2016)Yahata, Morimoto, Hashimoto, Lisi, Shibata,
  Kawakubo, Kuwabara, Kuroda, Yamada, Megumi, et~al.]{yahata2016small}
N.~Yahata, J.~Morimoto, R.~Hashimoto, G.~Lisi, K.~Shibata, Y.~Kawakubo,
  H.~Kuwabara, M.~Kuroda, T.~Yamada, F.~Megumi \emph{et~al.}, ``A small number
  of abnormal brain connections predicts adult autism spectrum disorder,''
  \emph{Nat. Commun.}, vol.~7, pp. 1--12, 2016.

\bibitem[Lindquist(2008)]{lindquist2008statistical}
M.~A. Lindquist, ``The statistical analysis of fmri data,'' \emph{Stat Sci},
  vol.~23, pp. 439--464, 2008.

\bibitem[Smith(2012)]{smith2012future}
S.~M. Smith, ``The future of fmri connectivity,'' \emph{NeuroImage}, vol.~62,
  pp. 1257--1266, 2012.

\bibitem[Shi and Guo(2016)]{shi2016}
R.~Shi and Y.~Guo, ``Investigating differences in brain functional networks
  using hierarchical covariate-adjusted independent component analysis,''
  \emph{Ann Appl Stat}, p. 1930, 2016.

\bibitem[Dai et~al.(2017)Dai, Guo, Initiative, et~al.]{dai2017predicting}
T.~Dai, Y.~Guo, A.~D.~N. Initiative \emph{et~al.}, ``Predicting individual
  brain functional connectivity using a bayesian hierarchical model,''
  \emph{NeuroImage}, vol. 147, pp. 772--787, 2017.

\bibitem[Higgins et~al.(2019)Higgins, Kundu, Choi, Mayberg, and
  Guo]{higgins2019difference}
I.~A. Higgins, S.~Kundu, K.~S. Choi, H.~S. Mayberg, and Y.~Guo, ``A difference
  degree test for comparing brain networks,'' \emph{Hum Brain Mapp}, pp.
  4518--4536, 2019.

\bibitem[Jie et~al.(2016)Jie, Liu, Jiang, and Zhang]{jie2016sub}
B.~Jie, M.~Liu, X.~Jiang, and D.~Zhang, ``Sub-network based kernels for brain
  network classification,'' in \emph{ICBC}, 2016.

\bibitem[He et~al.(2018)He, Chen, Xu, Zhou, and Wang]{he2018boosted}
L.~He, K.~Chen, W.~Xu, J.~Zhou, and F.~Wang, ``Boosted sparse and low-rank
  tensor regression,'' in \emph{NeurIPS}, 2018.

\bibitem[Liu et~al.(2018{\natexlab{b}})Liu, He, Cao, Yu, Ragin, and
  Leow]{liu2018multi}
Y.~Liu, L.~He, B.~Cao, P.~Yu, A.~Ragin, and A.~Leow, ``Multi-view multi-graph
  embedding for brain network clustering analysis,'' in \emph{AAAI}, 2018.

\bibitem[Schlichtkrull et~al.(2018)Schlichtkrull, Kipf, Bloem, Van Den~Berg,
  Titov, and Welling]{schlichtkrull2018modeling}
M.~Schlichtkrull, T.~N. Kipf, P.~Bloem, R.~Van Den~Berg, I.~Titov, and
  M.~Welling, ``Modeling relational data with graph convolutional networks,''
  in \emph{ESWC}, 2018.

\bibitem[Wu et~al.(2019)Wu, Tang, Zhu, Wang, Xie, and Tan]{wu2019session}
S.~Wu, Y.~Tang, Y.~Zhu, L.~Wang, X.~Xie, and T.~Tan, ``Session-based
  recommendation with graph neural networks,'' in \emph{AAAI}, 2019.

\bibitem[Li et~al.(2019)Li, Dvornek, Zhou, Zhuang, Ventola, and
  Duncan]{li2019graph}
X.~Li, N.~C. Dvornek, Y.~Zhou, J.~Zhuang, P.~Ventola, and J.~S. Duncan, ``Graph
  neural network for interpreting task-fmri biomarkers,'' in \emph{MICCAI},
  2019.

\bibitem[Bessadok et~al.(2021)Bessadok, Mahjoub, and Rekik]{bessadok2021graph}
A.~Bessadok, M.~A. Mahjoub, and I.~Rekik, ``Graph neural networks in network
  neuroscience,'' \emph{arXiv.org}, 2021.

\bibitem[Cui et~al.(2022{\natexlab{a}})Cui, Dai, Zhu, Li, He, and
  Yang]{cui2022interpretable}
H.~Cui, W.~Dai, Y.~Zhu, X.~Li, L.~He, and C.~Yang, ``Interpretable graph neural
  networks for connectome-based brain disorder analysis,'' in \emph{MICCAI},
  2022.

\bibitem[Zhu et~al.(2022)Zhu, Cui, He, Sun, and Yang]{zhu2022joint}
Y.~Zhu, H.~Cui, L.~He, L.~Sun, and C.~Yang, ``Joint embedding of structural and
  functional brain networks with graph neural networks for mental illness
  diagnosis,'' 2022.

\bibitem[Kan et~al.(2022{\natexlab{a}})Kan, Cui, Joshua, Guo, and
  Yang]{xuan2022fbnetgen}
X.~Kan, H.~Cui, L.~Joshua, Y.~Guo, and C.~Yang, ``Fbnetgen: Task-aware
  gnn-based fmri analysis via functional brain network generation,'' in
  \emph{MIDL}, 2022.

\bibitem[Tang et~al.(2022{\natexlab{a}})Tang, Guo, Fu, Qu, Thompson, Huang, and
  Zhan]{tang2022hierarchical2}
H.~Tang, L.~Guo, X.~Fu, B.~Qu, P.~M. Thompson, H.~Huang, and L.~Zhan,
  ``Hierarchical brain embedding using explainable graph learning,'' in
  \emph{2022 IEEE 19th International Symposium on Biomedical Imaging
  (ISBI)}.\hskip 1em plus 0.5em minus 0.4em\relax IEEE, 2022, pp. 1--5.

\bibitem[Tang et~al.(2022{\natexlab{b}})Tang, Guo, Fu, Qu, Ajilore, Wang,
  Thompson, Huang, Leow, and Zhan]{tang2022hierarchical}
H.~Tang, L.~Guo, X.~Fu, B.~Qu, O.~Ajilore, Y.~Wang, P.~M. Thompson, H.~Huang,
  A.~D. Leow, and L.~Zhan, ``A hierarchical graph learning model for brain
  network regression analysis,'' \emph{Frontiers in Neuroscience}, vol.~16,
  2022.

\bibitem[Sarraf and Sun(2016)]{sarraf2016functional}
S.~Sarraf and J.~Sun, ``Functional brain imaging: A comprehensive survey,''
  \emph{arXiv.org}, 2016.

\bibitem[Bernstein et~al.(2018)Bernstein, Akzhigitov, Kondrateva,
  Sushchinskaya, Samotaeva, and Gaskin]{DBLP:journals/mda/BernsteinAKSSG18}
A.~Bernstein, R.~Akzhigitov, E.~Kondrateva, S.~Sushchinskaya, I.~Samotaeva, and
  V.~Gaskin, ``{MRI} brain imagery processing software in data analysis,''
  \emph{Trans. Mass Data Anal. Images Signals}, vol.~9, pp. 3--17, 2018.

\bibitem[Ganis and Kosslyn(2002)]{GANIS2002493}
G.~Ganis and S.~M. Kosslyn, ``Neuroimaging,'' in \emph{Encyclopedia of the
  Human Brain}, 2002, pp. 493--505.

\bibitem[Lang et~al.(2012)Lang, Tom{\'{e}}, Keck, S{\'{a}}ez, and
  Puntonet]{DBLP:journals/cin/LangTKSP12}
E.~W. Lang, A.~M. Tom{\'{e}}, I.~R. Keck, J.~M.~G. S{\'{a}}ez, and C.~G.
  Puntonet, ``Brain connect analysis: {A} short survey,'' \emph{Comput. Intell.
  Neurosci.}, vol. 2012, pp. 412\,512:1--412\,512:21, 2012.

\bibitem[Bellec et~al.(2017)Bellec, Chu, Chouinard-Decorte, Benhajali,
  Margulies, and Craddock]{bellec2017neuro}
P.~Bellec, C.~Chu, F.~Chouinard-Decorte, Y.~Benhajali, D.~S. Margulies, and
  R.~C. Craddock, ``The neuro bureau adhd-200 preprocessed repository,''
  \emph{NeuroImage}, vol. 144, pp. 275--286, 2017.

\bibitem[Petersen et~al.(2010)Petersen, Aisen, Beckett, Donohue, Gamst, Harvey,
  Jack, Jagust, Shaw, Toga, et~al.]{petersen2010alzheimer}
R.~C. Petersen, P.~Aisen, L.~A. Beckett, M.~Donohue, A.~Gamst, D.~J. Harvey,
  C.~Jack, W.~Jagust, L.~Shaw, A.~Toga \emph{et~al.}, ``Alzheimer's disease
  neuroimaging initiative (adni): clinical characterization,''
  \emph{Neurology}, vol.~74, pp. 201--209, 2010.

\bibitem[Van~Essen et~al.(2012)Van~Essen, Ugurbil, Auerbach, Barch, Behrens,
  Bucholz, Chang, Chen, Corbetta, Curtiss, et~al.]{van2012human}
D.~C. Van~Essen, K.~Ugurbil, E.~Auerbach, D.~Barch, T.~E. Behrens, R.~Bucholz,
  A.~Chang, L.~Chen, M.~Corbetta, S.~W. Curtiss \emph{et~al.}, ``The human
  connectome project: a data acquisition perspective,'' \emph{NeuroImage},
  vol.~62, pp. 2222--2231, 2012.

\bibitem[Di~Martino et~al.(2014)Di~Martino, Yan, Li, Denio, Castellanos,
  Alaerts, Anderson, Assaf, Bookheimer, Dapretto, et~al.]{di2014autism}
A.~Di~Martino, C.-G. Yan, Q.~Li, E.~Denio, F.~X. Castellanos, K.~Alaerts, J.~S.
  Anderson, M.~Assaf, S.~Y. Bookheimer, M.~Dapretto \emph{et~al.}, ``The autism
  brain imaging data exchange: towards a large-scale evaluation of the
  intrinsic brain architecture in autism,'' \emph{Mol. Psychiatry}, vol.~19,
  pp. 659--667, 2014.

\bibitem[Wang et~al.(2016)Wang, Kang, Kemmer, and Guo]{wang2016efficient}
Y.~Wang, J.~Kang, P.~B. Kemmer, and Y.~Guo, ``An efficient and reliable
  statistical method for estimating functional connectivity in large scale
  brain networks using partial correlation,'' \emph{Front. Neurosci.}, vol.~10,
  p. 123, 2016.

\bibitem[Smith et~al.(2011)Smith, Miller, Salimi-Khorshidi, Webster, Beckmann,
  Nichols, Ramsey, and Woolrich]{smith2011network}
S.~M. Smith, K.~L. Miller, G.~Salimi-Khorshidi, M.~Webster, C.~F. Beckmann,
  T.~E. Nichols, J.~D. Ramsey, and M.~W. Woolrich, ``Network modelling methods
  for fmri,'' \emph{NeuroImage}, vol.~54, pp. 875--891, 2011.

\bibitem[Marek et~al.(2011)Marek, Jennings, Lasch, Siderowf, Tanner, Simuni,
  Coffey, Kieburtz, Flagg, Chowdhury, et~al.]{marek2011parkinson}
K.~Marek, D.~Jennings, S.~Lasch, A.~Siderowf, C.~Tanner, T.~Simuni, C.~Coffey,
  K.~Kieburtz, E.~Flagg, S.~Chowdhury \emph{et~al.}, ``The parkinson
  progression marker initiative (ppmi),'' \emph{Prog. Neurobiol.}, vol.~95, pp.
  629--635, 2011.

\bibitem[Ellis et~al.(2009)Ellis, Bush, Darby, De~Fazio, Foster, Hudson,
  Lautenschlager, Lenzo, Martins, Maruff, et~al.]{ellis2009australian}
K.~A. Ellis, A.~I. Bush, D.~Darby, D.~De~Fazio, J.~Foster, P.~Hudson, N.~T.
  Lautenschlager, N.~Lenzo, R.~N. Martins, P.~Maruff \emph{et~al.}, ``The
  australian imaging, biomarkers and lifestyle (aibl) study of aging:
  methodology and baseline characteristics of 1112 individuals recruited for a
  longitudinal study of alzheimer's disease,'' \emph{Int Psychogeriatr},
  vol.~21, pp. 672--687, 2009.

\bibitem[Leow et~al.(2009)Leow, Zhu, Zhan, McMahon, de~Zubicaray, Meredith,
  Wright, Toga, and Thompson]{leow2009tensor}
A.~D. Leow, S.~Zhu, L.~Zhan, K.~McMahon, G.~I. de~Zubicaray, M.~Meredith,
  M.~Wright, A.~Toga, and P.~Thompson, ``The tensor distribution function,''
  \emph{Magn Reson Med}, vol.~61, pp. 205--214, 2009.

\bibitem[Behrens et~al.(2003)Behrens, Woolrich, Jenkinson, Johansen-Berg,
  Nunes, Clare, Matthews, Brady, and Smith]{behrens2003characterization}
T.~E. Behrens, M.~W. Woolrich, M.~Jenkinson, H.~Johansen-Berg, R.~G. Nunes,
  S.~Clare, P.~M. Matthews, J.~M. Brady, and S.~M. Smith, ``Characterization
  and propagation of uncertainty in diffusion-weighted mr imaging,'' \emph{Magn
  Reson Med}, vol.~50, pp. 1077--1088, 2003.

\bibitem[Basser et~al.(2000)Basser, Pajevic, Pierpaoli, Duda, and
  Aldroubi]{basser2000vivo}
P.~J. Basser, S.~Pajevic, C.~Pierpaoli, J.~Duda, and A.~Aldroubi, ``In vivo
  fiber tractography using dt-mri data,'' \emph{Magn Reson Med}, vol.~44, pp.
  625--632, 2000.

\bibitem[Behrens et~al.(2007)Behrens, Berg, Jbabdi, Rushworth, and
  Woolrich]{behrens2007probabilistic}
T.~E. Behrens, H.~J. Berg, S.~Jbabdi, M.~F. Rushworth, and M.~W. Woolrich,
  ``Probabilistic diffusion tractography with multiple fibre orientations: What
  can we gain?'' \emph{NeuroImage}, vol.~34, pp. 144--155, 2007.

\bibitem[Zhan et~al.(2015{\natexlab{b}})Zhan, Zhou, Wang, Jin, Jahanshad,
  Prasad, Nir, Leonardo, Ye, Thompson, et~al.]{zhan2015comparison}
L.~Zhan, J.~Zhou, Y.~Wang, Y.~Jin, N.~Jahanshad, G.~Prasad, T.~M. Nir, C.~D.
  Leonardo, J.~Ye, P.~M. Thompson \emph{et~al.}, ``Comparison of nine
  tractography algorithms for detecting abnormal structural brain networks in
  alzheimer’s disease,'' \emph{Front. Aging Neurosci.}, vol.~7, p.~48, 2015.

\bibitem[Deshpande and Jia(2020)]{deshpande2020multi}
G.~Deshpande and H.~Jia, ``Multi-level clustering of dynamic directional brain
  network patterns and their behavioral relevance,'' \emph{Front. Neurosci.},
  vol.~13, p. 1448, 2020.

\bibitem[Zhu et~al.(2019)Zhu, Li, Huang, Xu, Guan, and Zhang]{zhu2019hybrid}
Q.~Zhu, H.~Li, J.~Huang, X.~Xu, D.~Guan, and D.~Zhang, ``Hybrid functional
  brain network with first-order and second-order information for
  computer-aided diagnosis of schizophrenia,'' \emph{Front. Neurosci.}, p. 603,
  2019.

\bibitem[Espinoza et~al.(2019)Espinoza, Vergara, Damaraju, Henke, Faghiri,
  Turner, Belger, Ford, McEwen, Mathalon, et~al.]{espinoza2019characterizing}
F.~A. Espinoza, V.~M. Vergara, E.~Damaraju, K.~G. Henke, A.~Faghiri, J.~A.
  Turner, A.~A. Belger, J.~M. Ford, S.~C. McEwen, D.~H. Mathalon \emph{et~al.},
  ``Characterizing whole brain temporal variation of functional connectivity
  via zero and first order derivatives of sliding window correlations,''
  \emph{Front. Neurosci.}, vol.~13, p. 634, 2019.

\bibitem[Kuang et~al.(2019)Kuang, Zhao, Xing, Chen, Xiong, and
  Han]{kuang2019metabolic}
L.~Kuang, D.~Zhao, J.~Xing, Z.~Chen, F.~Xiong, and X.~Han, ``Metabolic brain
  network analysis of fdg-pet in alzheimer’s disease using kernel-based
  persistent features,'' \emph{Molecules}, vol.~24, p. 2301, 2019.

\bibitem[Joudaki et~al.(2012)Joudaki, Salehi, Jalili, and
  Knyazeva]{joudaki2012eeg}
A.~Joudaki, N.~Salehi, M.~Jalili, and M.~G. Knyazeva, ``Eeg-based functional
  brain networks: does the network size matter?'' \emph{PLoS One}, vol.~7, p.
  e35673, 2012.

\bibitem[Calhoun and Sui(2016)]{calhoun2016multimodal}
V.~D. Calhoun and J.~Sui, ``Multimodal fusion of brain imaging data: a key to
  finding the missing link (s) in complex mental illness,'' \emph{Biol
  Psychiatry Cogn Neurosci Neuroimaging}, vol.~1, pp. 230--244, 2016.

\bibitem[Cui et~al.(2022{\natexlab{b}})Cui, Lu, Li, and
  Yang]{cui2021positional}
H.~Cui, Z.~Lu, P.~Li, and C.~Yang, ``On positional and structural node features
  for graph neural networks on non-attributed graphs,'' \emph{CIKM}, 2022.

\bibitem[Duong et~al.(2019)Duong, Hoang, Dang, Nguyen, and
  Aberer]{duong2019node}
C.~T. Duong, T.~D. Hoang, H.~T.~H. Dang, Q.~V.~H. Nguyen, and K.~Aberer, ``On
  node features for graph neural networks,'' \emph{arXiv.org}, 2019.

\bibitem[Errica et~al.(2020)Errica, Podda, Bacciu, and
  Micheli]{errica_fair_2020}
F.~Errica, M.~Podda, D.~Bacciu, and A.~Micheli, ``A fair comparison of graph
  neural networks for graph classification,'' in \emph{ICLR}, 2020.

\bibitem[You et~al.(2019)You, Ying, and Leskovec]{you2019position}
J.~You, R.~Ying, and J.~Leskovec, ``Position-aware graph neural networks,'' in
  \emph{ICML}, 2019.

\bibitem[Huang et~al.(2021)Huang, He, Singh, Lim, and
  Benson]{DBLP:journals/corr/abs-2010-13993}
Q.~Huang, H.~He, A.~Singh, S.~Lim, and A.~R. Benson, ``Combining label
  propagation and simple models out-performs graph neural networks,'' in
  \emph{ICLR}, 2021.

\bibitem[Chaudhuri et~al.(2012)Chaudhuri, Chung, and
  Tsiatas]{chaudhuri2012spectral}
K.~Chaudhuri, F.~Chung, and A.~Tsiatas, ``Spectral clustering of graphs with
  general degrees in the extended planted partition model,'' in \emph{COLT},
  2012.

\bibitem[Zhang and Rohe(2018)]{DBLP:conf/nips/ZhangR18}
Y.~Zhang and K.~Rohe, ``Understanding regularized spectral clustering via graph
  conductance,'' in \emph{NeurIPS}, 2018.

\bibitem[Cai and Wang(2018)]{Cai:2018te}
C.~Cai and Y.~Wang, ``{A Simple Yet Effective Baseline for Non-Attributed Graph
  Classification},'' \emph{arXiv.org}, 2018.

\bibitem[Vaswani et~al.(2017)Vaswani, Shazeer, Parmar, Uszkoreit, Jones, Gomez,
  Kaiser, and Polosukhin]{vaswani2017attention}
A.~Vaswani, N.~Shazeer, N.~Parmar, J.~Uszkoreit, L.~Jones, A.~N. Gomez,
  {\L}.~Kaiser, and I.~Polosukhin, ``Attention is all you need,''
  \emph{NeurIPS}, 2017.

\bibitem[Niu et~al.(2021)Niu, Zhong, and Yu]{niu2021review}
Z.~Niu, G.~Zhong, and H.~Yu, ``A review on the attention mechanism of deep
  learning,'' \emph{Neurocomputing}, vol. 452, pp. 48--62, 2021.

\bibitem[Devlin et~al.(2019)Devlin, Chang, Lee, and
  Toutanova]{devlin-etal-2019-bert}
J.~Devlin, M.-W. Chang, K.~Lee, and K.~Toutanova, ``{BERT}: Pre-training of
  deep bidirectional transformers for language understanding,'' in \emph{ACL},
  2019.

\bibitem[Guo et~al.(2021)Guo, Xu, Liu, Liu, Jiang, Mu, Zhang, Martin, Cheng,
  and Hu]{guo2021attention_survey}
M.-H. Guo, T.-X. Xu, J.-J. Liu, Z.-N. Liu, P.-T. Jiang, T.-J. Mu, S.-H. Zhang,
  R.~R. Martin, M.-M. Cheng, and S.-M. Hu, ``Attention mechanisms in computer
  vision: A survey,'' \emph{arXiv.org}, 2021.

\bibitem[Grattarola et~al.(2021)Grattarola, Zambon, Bianchi, and
  Alippi]{grattarola2021understanding}
D.~Grattarola, D.~Zambon, F.~M. Bianchi, and C.~Alippi, ``Understanding pooling
  in graph neural networks,'' \emph{arXiv.org}, 2021.

\bibitem[Mesquita et~al.(2020)Mesquita, Souza, and
  Kaski]{mesquita2020rethinking}
D.~Mesquita, A.~Souza, and S.~Kaski, ``Rethinking pooling in graph neural
  networks,'' \emph{NeurIPS}, 2020.

\bibitem[Ying et~al.(2018)Ying, You, Morris, Ren, Hamilton, and
  Leskovec]{ying2018hierarchical}
Z.~Ying, J.~You, C.~Morris, X.~Ren, W.~Hamilton, and J.~Leskovec,
  ``Hierarchical graph representation learning with differentiable pooling,''
  in \emph{NeurIPS}, 2018.

\bibitem[Gopinath et~al.(2020)Gopinath, Desrosiers, and
  Lombaert]{gopinath2020learnable}
K.~Gopinath, C.~Desrosiers, and H.~Lombaert, ``Learnable pooling in graph
  convolution networks for brain surface analysis,'' \emph{IEEE Trans. Pattern
  Anal. Mach. Intell.}, 2020.

\bibitem[Kan et~al.(2022{\natexlab{b}})Kan, Dai, Cui, Zhang, Guo, and
  Carl]{kan2022transformer}
X.~Kan, W.~Dai, H.~Cui, Z.~Zhang, Y.~Guo, and Y.~Carl, ``Brain network
  transformer,'' in \emph{NeurIPS}, 2022.

\bibitem[Chen et~al.(2020)Chen, Beer, Tustison, Cook, Shinohara, Shou,
  Initiative, et~al.]{chen2020removal}
A.~A. Chen, J.~C. Beer, N.~J. Tustison, P.~A. Cook, R.~T. Shinohara, H.~Shou,
  A.~D.~N. Initiative \emph{et~al.}, ``Removal of scanner effects in covariance
  improves multivariate pattern analysis in neuroimaging data,''
  \emph{bioRxiv}, p. 858415, 2020.

\bibitem[Fortin et~al.(2018)Fortin, Cullen, Sheline, Taylor, Aselcioglu, Cook,
  Adams, Cooper, Fava, McGrath, et~al.]{fortin2018harmonization}
J.-P. Fortin, N.~Cullen, Y.~I. Sheline, W.~D. Taylor, I.~Aselcioglu, P.~A.
  Cook, P.~Adams, C.~Cooper, M.~Fava, P.~J. McGrath \emph{et~al.},
  ``Harmonization of cortical thickness measurements across scanners and
  sites,'' \emph{NeuroImage}, vol. 167, pp. 104--120, 2018.

\bibitem[Chen et~al.(2014)Chen, Liu, Calhoun, Arias-Vasquez, Zwiers, Gupta,
  Franke, and Turner]{chen2014exploration}
J.~Chen, J.~Liu, V.~D. Calhoun, A.~Arias-Vasquez, M.~P. Zwiers, C.~N. Gupta,
  B.~Franke, and J.~A. Turner, ``Exploration of scanning effects in multi-site
  structural mri studies,'' \emph{J. Neurosci. Methods}, vol. 230, pp. 37--50,
  2014.

\bibitem[Bell et~al.(2022)Bell, Godfrey, Ware, Yeates, and
  Harris]{bell2022harmonization}
T.~K. Bell, K.~J. Godfrey, A.~L. Ware, K.~O. Yeates, and A.~D. Harris,
  ``Harmonization of multi-site mrs data with combat,'' \emph{NeuroImage}, p.
  119330, 2022.

\bibitem[Pomponio et~al.(2020)Pomponio, Erus, Habes, Doshi, Srinivasan,
  Mamourian, Bashyam, Nasrallah, Satterthwaite, Fan,
  et~al.]{pomponio2020harmonization}
R.~Pomponio, G.~Erus, M.~Habes, J.~Doshi, D.~Srinivasan, E.~Mamourian,
  V.~Bashyam, I.~M. Nasrallah, T.~D. Satterthwaite, Y.~Fan \emph{et~al.},
  ``Harmonization of large mri datasets for the analysis of brain imaging
  patterns throughout the lifespan,'' \emph{NeuroImage}, vol. 208, p. 116450,
  2020.

\bibitem[Yamashita et~al.(2019)Yamashita, Yahata, Itahashi, Lisi, Yamada,
  Ichikawa, Takamura, Yoshihara, Kunimatsu, Okada,
  et~al.]{yamashita2019harmonization}
A.~Yamashita, N.~Yahata, T.~Itahashi, G.~Lisi, T.~Yamada, N.~Ichikawa,
  M.~Takamura, Y.~Yoshihara, A.~Kunimatsu, N.~Okada \emph{et~al.},
  ``Harmonization of resting-state functional mri data across multiple imaging
  sites via the separation of site differences into sampling bias and
  measurement bias,'' \emph{PLOS Biol.}, vol.~17, p. e3000042, 2019.

\bibitem[Pinto et~al.(2020)Pinto, Paolella, Billiet, Van~Dyck, Guns, Jeurissen,
  Ribbens, den Dekker, and Sijbers]{pinto2020harmonization}
M.~S. Pinto, R.~Paolella, T.~Billiet, P.~Van~Dyck, P.-J. Guns, B.~Jeurissen,
  A.~Ribbens, A.~J. den Dekker, and J.~Sijbers, ``Harmonization of brain
  diffusion mri: Concepts and methods,'' \emph{Front. Neurosci.}, vol.~14, p.
  396, 2020.

\bibitem[Tzourio-Mazoyer et~al.(2002)Tzourio-Mazoyer, Landeau, Papathanassiou,
  Crivello, Etard, Delcroix, Mazoyer, and Joliot]{tzourio2002automated}
N.~Tzourio-Mazoyer, B.~Landeau, D.~Papathanassiou, F.~Crivello, O.~Etard,
  N.~Delcroix, B.~Mazoyer, and M.~Joliot, ``Automated anatomical labeling of
  activations in spm using a macroscopic anatomical parcellation of the mni mri
  single-subject brain,'' \emph{NeuroImage}, vol.~15, pp. 273--289, 2002.

\bibitem[Power et~al.(2011)Power, Cohen, Nelson, Wig, Barnes, Church, Vogel,
  Laumann, Miezin, Schlaggar, et~al.]{power264}
J.~D. Power, A.~L. Cohen, S.~M. Nelson, G.~S. Wig, K.~A. Barnes, J.~A. Church,
  A.~C. Vogel, T.~O. Laumann, F.~M. Miezin, B.~L. Schlaggar \emph{et~al.},
  ``{Functional Network Organization of the Human Brain},'' \emph{Neuron},
  vol.~72, pp. 665--678, 2011.

\bibitem[Smith et~al.(2009)Smith, Fox, Miller, Glahn, Fox, Mackay, Filippini,
  Watkins, Toro, Laird, et~al.]{smith2009correspondence}
S.~M. Smith, P.~T. Fox, K.~L. Miller, D.~C. Glahn, P.~M. Fox, C.~E. Mackay,
  N.~Filippini, K.~E. Watkins, R.~Toro, A.~R. Laird \emph{et~al.},
  ``Correspondence of the brain's functional architecture during activation and
  rest,'' \emph{Proc. Natl. Acad. Sci. U.S.A.}, vol. 106, pp. 13\,040--13\,045,
  2009.

\bibitem[Desikan et~al.(2006)Desikan, S{\'e}gonne, Fischl, Quinn, Dickerson,
  Blacker, Buckner, Dale, Maguire, Hyman, et~al.]{desikan2006automated}
R.~S. Desikan, F.~S{\'e}gonne, B.~Fischl, B.~T. Quinn, B.~C. Dickerson,
  D.~Blacker, R.~L. Buckner, A.~M. Dale, R.~P. Maguire, B.~T. Hyman
  \emph{et~al.}, ``An automated labeling system for subdividing the human
  cerebral cortex on mri scans into gyral based regions of interest,''
  \emph{NeuroImage}, vol.~31, pp. 968--980, 2006.

\bibitem[Casey et~al.(2018)Casey, Cannonier, Conley, Cohen, Barch, Heitzeg,
  Soules, Teslovich, Dellarco, Garavan, et~al.]{casey2018adolescent}
B.~Casey, T.~Cannonier, M.~I. Conley, A.~O. Cohen, D.~M. Barch, M.~M. Heitzeg,
  M.~E. Soules, T.~Teslovich, D.~V. Dellarco, H.~Garavan \emph{et~al.}, ``The
  adolescent brain cognitive development (abcd) study: imaging acquisition
  across 21 sites,'' \emph{Dev Cogn Neurosci}, vol.~32, pp. 43--54, 2018.

\bibitem[Glasser et~al.(2013)Glasser, Sotiropoulos, Wilson, Coalson, Fischl,
  Andersson, Xu, Jbabdi, Webster, Polimeni, {Van Essen}, and
  Jenkinson]{GLASSER2013105}
M.~F. Glasser, S.~N. Sotiropoulos, J.~A. Wilson, T.~S. Coalson, B.~Fischl,
  J.~L. Andersson, J.~Xu, S.~Jbabdi, M.~Webster, J.~R. Polimeni, D.~C. {Van
  Essen}, and M.~Jenkinson, ``The minimal preprocessing pipelines for the human
  connectome project,'' \emph{NeuroImage}, vol.~80, pp. 105--124, 2013.

\bibitem[Lu et~al.(2008)]{Lu:2008cw}
H.~Lu \emph{et~al.}, ``Mpca: Multilinear principal component analysis of tensor
  objects,'' \emph{IEEE Trans. Neural Netw.}, 2008.

\bibitem[Dyrba et~al.(2015)]{Dyrba:2015ci}
M.~Dyrba \emph{et~al.}, ``Multimodal analysis of functional and structural
  disconnection in alzheimer's disease using multiple kernel svm,'' \emph{Hum
  Brain Mapp}, 2015.

\bibitem[Nie et~al.(2011)Nie, Zeng, Tsang, Xu, and Zhang]{nie2011spectral}
F.~Nie, Z.~Zeng, I.~W. Tsang, D.~Xu, and C.~Zhang, ``Spectral embedded
  clustering: A framework for in-sample and out-of-sample spectral
  clustering,'' \emph{IEEE Trans. Neural Netw.}, vol.~22, pp. 1796--1808, 2011.

\bibitem[Nie et~al.(2016)Nie, Li, Li, et~al.]{nie2016parameter}
F.~Nie, J.~Li, X.~Li \emph{et~al.}, ``Parameter-free auto-weighted multiple
  graph learning: a framework for multiview clustering and semi-supervised
  classification.'' in \emph{IJCAI}, 2016.

\bibitem[Paszke et~al.(2019)Paszke, Gross, et~al.]{Paszke:2019vf}
A.~Paszke, S.~Gross \emph{et~al.}, ``{PyTorch: an imperative style,
  high-performance deep learning library},'' in \emph{NeurIPS}, 2019.

\bibitem[Fey and Lenssen(2019)]{Fey:2019wv}
M.~Fey and J.~E. Lenssen, ``{Fast graph representation learning with PyTorch
  Geometric},'' in \emph{RLGM@ICLR}, 2019.

\bibitem[Yang et~al.(2022)Yang, Zhu, Cui, Kan, He, Guo, and Yang]{yang2022data}
Y.~Yang, Y.~Zhu, H.~Cui, X.~Kan, L.~He, Y.~Guo, and C.~Yang, ``Data-efficient
  brain connectome analysis via multi-task meta-learning,'' \emph{KDD}, 2022.

\end{thebibliography}

\end{document}